\documentclass[aps,prb,reprint,groupedaddress,floatfix]{revtex4-1}

\usepackage[english]{babel} 
\usepackage{amssymb,amsmath}
\usepackage{graphicx}
\usepackage{url}
\usepackage{multirow}
\usepackage{chemformula}
\usepackage{tabularx}

\newlength{\smallpic}
\begin{document}

\title{Computational exfoliation of atomically thin 1D materials with application to Majorana bound states}
\author{Hadeel Moustafa, Peter Mahler Larsen, Morten N. Gjerding, Jens Jørgen Mortensen, Kristian S. Thygesen, Karsten W. Jacobsen
 }
\date{\today}

\begin{abstract}
 We introduce a computational database with calculated structural, thermodynamic, electronic, magnetic, and optical properties of 820 one-dimensional materials. The materials are systematically selected and exfoliated from experimental databases of crystal structures based on a dimensionality scoring parameter. The database is furthermore expanded by chemical element substitution in the materials. The materials are investigated in both their bulk form and as isolated one-dimensional components. We discuss the methodology behind the database, give an overview of some of the calculated properties, and look at patterns and correlations in the data. The database is furthermore applied in computational screening to identify materials, which could exhibit Majorana bound states.
\end{abstract}

\maketitle

\section{Introduction}

Low-dimensional materials are of interest because of their unique physical and chemical properties and because of several potential applications from light-absorbers over single-photon emitters to catalysts \cite{spiece2019nanoscale, yang2016tuning, Cheon:2017kn}. This has led to intense experimental and theoretical studies of, in particular, two-dimensional materials, where a number of experimental of techniques have been developed to produce the materials and investigate their properties \cite{zhou20192dmatpedia, singh2015computational}. 

One-dimensional materials are interesting as well but have been less explored, and techniques for their production are less developed. The reduced dimensionality gives rise to modified band structures, charge screening, and electron-phonon coupling paving the way for new material properties \cite{guo_one-dimensional_2022, balandin_one-dimensional_2022}. If sufficient control of the atomic and electronic structure can be achieved, it might be possible to investigate fundamental physical phenomena like Luttinger liquid behavior \cite{Haldane.1981, Bruus} and the presence of Majorana bound states \cite{Wilczek.2009, majorana}. Furthermore a number of potential applications have been suggested including photonic crystals \cite{jiang2004properties}, batteries \cite{tiwari2012zero}, transistors \cite{randle2018gate} and as electronic interconnects \cite{stolyarov2016breakdown,xia2003one}. In heterogeneous catalysis, it is well-known that low-dimensional structures like step edges can be particularly reactive \cite{Dahl.1999}, because the active step sites are electronically and geometrically different from the sites on planar surfaces. One-dimensional materials may therefore also be expected to exhibit special catalytic properties. Recent studies point to interesting methanol oxidation activity of alloy nanochains \cite{jiang_synergism_2020}.

The individual layers in a bulk two-dimensional material may exhibit different properties from the bulk material. For example, a monolayer of \ch{MoS2} has a direct band gap, while the band gap of bulk \ch{MoS2} is indirect. The stacking of different two-dimensional materials into heterostructures opens up even wider possibilities for materials design \cite{Geim.2013, Li.2020qjn}.  Likewise, one-dimensional materials can be combined with other one- or two-dimensional materials into new van der Waals heterostructures of mixed dimensionality allowing for tailored physical or chemical properties, which cannot be obtained by the individual components alone \cite{jariwala2017mixed, Liu.2019w4, gao2018generalized}.

The large interest in low-dimensional materials has given rise to the establishment of a few databases with computed material properties, in particular for two-dimensional materials \cite{Mounet:2018ks, Cheon:2017kn, Haastrup:2018ca}. The databases vary in their scope and in the range of material properties investigated.

In this paper, we introduce a database with calculated structural and electronic properties of specifically one-dimensional materials. One characteristic of the database is the systematic approach to the selection and characterization of the materials. The database contains two sets of materials. The first set, which we shall refer to as the \emph{core} of the database, consists of materials and material components identified in the inorganic crystal structure database (ICSD) \cite{Bergerhoff:1983hj} and the crystallography open database (COD) \cite{Grazulis:2011if}. These materials have been previously experimentally synthesized in their three-dimensional structure. The materials are selected using a recently developed dimensionality scoring parameter \cite{Larsen:2019cf}, which is based exclusively on the atomic geometry. The second set of materials, which we shall refer to as the \emph{shell}, are derived from the first set by chemical element substitution, where, say, a \ch{Ni} atom is substituted by another chemical element, like \ch{Pd} or \ch{Pt} with similar chemical properties. The element replaceability is taken from the statistical analysis by Glawe et al. \cite{Glawe:2016jg}.

As an illustration we use the database to identify materials which potentially could exhibit Majorana bound states. The materials are breaking inversion symmetry and further identified from the spin-orbit character of their band structure.

The resulting database, which we term C1DB, is available as part of the Computational Materials Repository \cite{c1db}.

\section{Materials selection}

The dimensionality of a bulk material and to which extent the material can be regarded as consisting of components of lower dimensionality is of course not rigorously well-defined. Depending on the property of interest, this being electronic, optical, mechanical or chemical, the material may exhibit different degrees of anisotropy, which can be interpreted as related to the dimensionality of the material. Another perspective on the dimensionality is whether a low-dimensional component of a material can be extracted from the material and stabilized in other environments outside the material. 
Here we shall follow Larsen et al.\ \cite{Larsen:2019cf} and take a simple geometric approach, where the dimensionality of a material and its components are determined exclusively based on interatomic distances. The first step in such an analysis is to determine which atoms in the solid are connected by bonds. We say that a bond between two atoms, $i$ and $j$, exists if the distance between the atoms $d_{ij}$ obeys
\begin{equation}
    d_{ij} < k(r_i^\text{cov}+r_j^\text{cov}),\label{eq:bonding}
\end{equation} 
where $r_i^\text{cov}$ and $r_j^\text{cov}$ are the covalent radii of the two atoms and $k$ is a parameter to be discussed further below. This is the same criterion used by Ashton et al. \cite{Ashton:2017hk} in their studies of layered materials, while Mounet et al. \cite{Mounet:2018ks} and Cheon et al.\ \cite{Cheon:2017kn} apply a slightly different criterion with an additive constant. 

Given the interatomic bonds, the dimension of a connected component can be determined as the rank of the subspace spanned by an atom and its periodically connected neighbors \cite{Mounet:2018ks,Larsen:2019cf}. It is possible to assess the dimensionality estimation by considering a few different values of the parameter $k$. However, it is informative to consider $k$ as a continuous parameter and use this to define a dimensionality scoring parameter \cite{Larsen:2019cf}.

For small values of $k$ in Eq.~\ref{eq:bonding} no bonds exist between the atoms, and the material thus consists of a collection of zero-dimensional components (i.e. the atoms). As $k$ is increased bonds begin to set in and for a particular value, $k_1$, one-dimensional components may form. At a higher value $k_2$ the one-dimensional components disappear and only two- or three-dimensional components are left. Not all materials exhibit a one-dimensional "phase", but jumps directly from dimension zero to dimension two or three corresponding to $k_2 = k_1$.

\begin{figure}[th]
    \centering
    \includegraphics[width=0.8\linewidth]{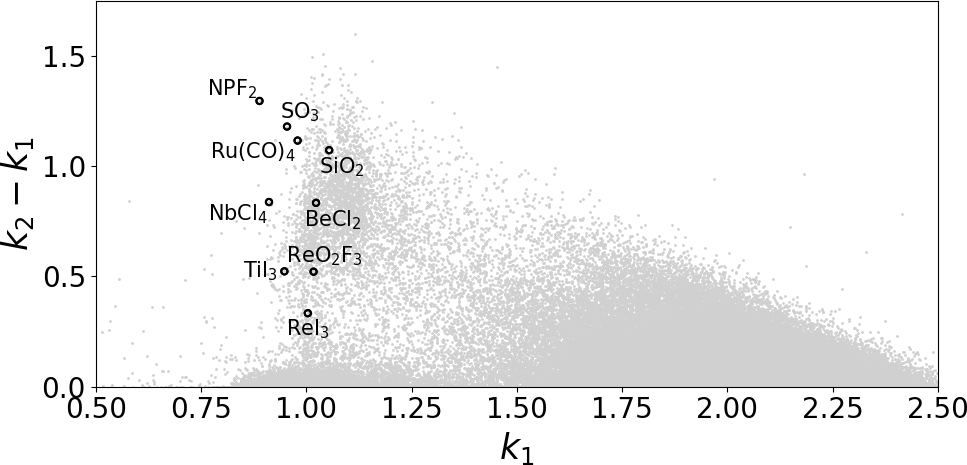}\\
    \includegraphics[width=0.8\linewidth]{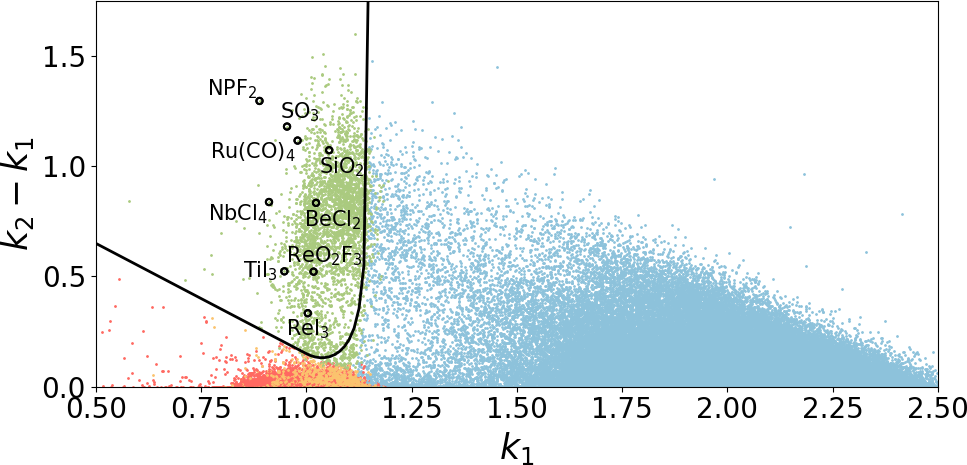}
    \caption{The $(k_1, k_2)$-values for dimension one for all the materials in ICSD and COD. A clear cluster representing the one-dimensional materials is seen in the upper figure. In the lower figure the line $s_{1D}=0.5$ is shown and the materials are colored according to their dimensionalities, where the dimension is determined as the one with the highest scoring parameter. One-dimensional materials are green, while \mbox{zero-,} two-, and three-dimensional materials are shown in blue, yellow, and red, respectively. (The figure is from Ref.~\onlinecite{Larsen:2019cf}.)}
    \label{fig:k-plot}
\end{figure}

Figure~\ref{fig:k-plot} shows the $k_1$ and $k_2$ values for the materials of ICSD and COD with some examples of potential one-dimensional materials indicated. A clear cluster representing the one-dimensional materials can be identified, and the materials are characterized by a high value of $k_2-k_1$ and a $k_1$-value not much larger than one. This allows for the definition of a scoring parameter \cite{Larsen:2019cf}, $s_1$,  as $s_1(k_1,k_2)=f(k_2)-f(k_1)$, where $f(x)=\tanh(c\cdot\max(0,x-1))$. High values of $s_1$ (i.e. values close to one) indicate a high degree of likelihood that the material is one-dimensional. Similar scoring parameters can be defined for the other dimensionalities, and by construction, the sum of the scoring values for the different dimensions add up to one. The line corresponding to $s_1=0.5$ is shown in the lower part of Figure~\ref{fig:k-plot}, and the materials where $s_1$ is the largest of the dimensional scoring parameters are indicated in green.

\begin{table}
    \centering
    \begin{tabular}{|l|r|}
    \hline
        Criterion&Number of materials\\ \hline
        Initial CMR database & 167767\\
        High 1D scoring parameter $s_1$& 3285\\
        $\le 80$ atoms in unit cell&1310\\
        $\le 4$ different chemical elements & 663\\
        $\le 20$ atoms in 1D components& 419\\
        Without missing atoms and other issues& 288\\
    \hline
    \end{tabular}
    \caption{Selection criteria for the 1D materials in the core of the database with the number of materials indicated.}
    \label{tab:selection}
\end{table}

The calculated dimensionality scoring parameters for the materials in ICSD and COD are available online at the Computational Materials Repository \cite{lowdim_database}, and this database is the starting point for our selection of core materials. 

\subsection{The materials in the core of the database}
We apply a number of selection criteria to identify the relevant materials. We require that the one-dimensional scoring parameter $s_1$ is greater than the scoring parameters for the other dimensionalities $s_0$, $s_2$, and $s_3$, and also greater than scoring parameters for combinations of dimensionalities like $s_{02}$, where both zero- and two-dimensional components are present. We consider only materials with 80 or less atoms in the unit cell and not more than 4 different chemical elements. We further limit ourselves to materials where the 1D-components have 20 atoms or less. Finally, we remove by hand a fairly large number of entries with various issues and inconsistencies. The ICSD and COD databases have a number of problematic entries, and because we focus on low-dimensional materials, we find a relative over-representation of entries with for example missing atoms. We remove entries with invalid structures, partial structures, theoretical structures, and missing atoms (in particular hydrogen). Finally, we also remove materials containing uranium, tantalum and technetium, because GPAW, the DFT code used in this work, does not have a PAW dataset for those elements. The number of materials identified at the different stages of the selection process is shown in Table~\ref{tab:selection}.
The selection leaves us with 288 materials, which constitute the core of the database. Three of these materials have only one chemical element present, 95 have two elements, 135 have three, and 55 have four elements.

The one-dimensional components of the materials are extracted using the approach described in Ref.~\onlinecite{Larsen:2019cf}. The code for performing the extraction is available as the module \textit{ase.geometry.dimensionality} in the Atomic Simulation Environment \cite{Larsen:2017hn}. Both the one-dimensional components and the original three-dimensional structures are kept in the database. The one-dimensional components are embedded in tetragonal unit cells with the $z$-axis in the direction of the components and with 8 Å of vacuum surrounding the structure.

\subsection{The materials in the shell: chemical element substitution}
The materials in the core of the database all originate from ICSD or COD, which means that they have been experimentally synthesized in their bulk form. We shall now expand the database with a \emph{shell} of potentially new materials obtained from the core by substitution of chemical elements.

It is well-known that some chemical elements have similar chemical properties. This is for example the case for elements in the same group of the periodic table.
In a recent study Glawe et al. \cite{Glawe:2016jg} investigate the correlations in the appearance of different chemical elements in the same crystal structure in the ICSD. They introduce a probability measure $P_{AB}$ which expresses how likely it is that if element A appears in a compound with a particular structure in ICSD, then the compound with B substituted for A in the same crystal structure will also be present in ICSD. We refer the reader to Ref.~\onlinecite{Glawe:2016jg} for the detailed construction of $P_{AB}$. 
We use the probability measure to create new entries in the database. By performing replacements with a high probability measure, the probability that the resulting material will be stable is increased. More specifically, we perform the substitutions $\ch{A} \rightarrow \ch{B}$ for elements A and B, where $P_{AB} > 0.2$. The probability measure is symmetric $P_{AB} = P_{BA}$, so if the substitution $\ch{A} \rightarrow \ch{B}$ is performed so is $\ch{B} \rightarrow \ch{A}$. The criterion $P_{AB} > 0.2$ gives rise to the following set of replacements: (S,Se,Te), (Br,Cl,I), (P,As), (H,F), (Mo,W), (Si,Ge), (Xe,Kr), (Cs,K,Rb), and (Ta,Nb). We perform the substitutions in materials with up to three different elements, and for a given core material we perform up to three element substitutions at a time. If a certain element is substituted then all the atoms of that element in the material are replaced. The substitutions are performed both in the bulk materials and in the one-dimensional components, so that the exfoliation energy for the new material can be calculated.
The replacement approach can of course be expanded to materials with more different elements and to multiple simultaneous substitutions of individual atoms. The present substitutions give rise to 532 new materials in the database, which we ascribe to the shell.

\begin{figure}
\centering
\includegraphics[width=1.0\linewidth]{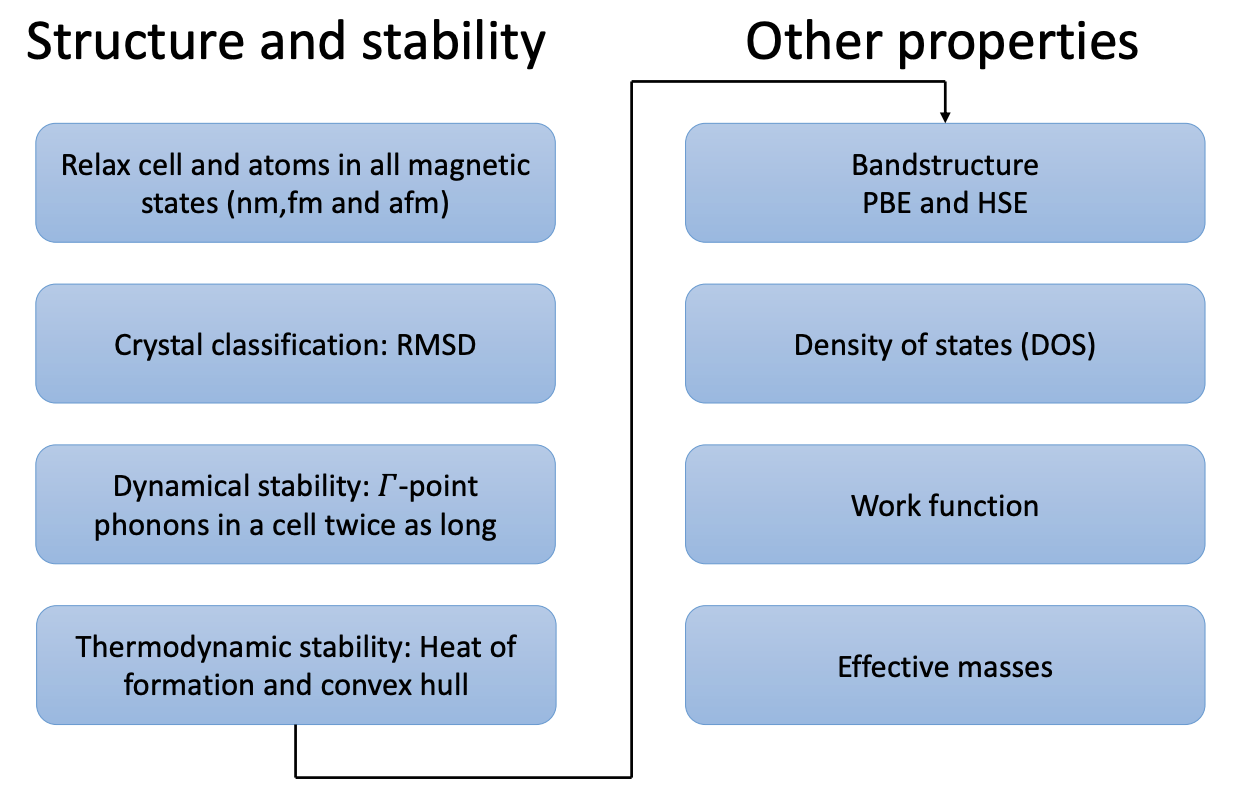}
\caption{A simple diagram of the workflow that is used to calculate the structure and properties of 1D materials. This is the workflow used to generate the data in our database. The first part of the workflow considers stability while the other part of our workflow evaluates different properties. A more detailed description of the steps in the workflow can be found in the text.
}
\label{fig:workflow}
\end{figure}

\section{Computational methods}
All electronic structure calculations are performed using density functional theory with the GPAW code \cite{Mortensen:2005ep,Enkovaara:2010jd} and the Atomic Simulation Environment \cite{Larsen:2017hn}. We employ a newly developed set of python modules entitled the Atomic Simulation Recipes (ASR) \cite{Gjerding.2021} with workflow management using MyQueue \cite{MyQueue, mortensen2020myqueue}.

ASR is a Python package that defines a set of basic common operations for handling simulations of atomic systems known as recipes. The 1D workflow has been using recipes for relaxation, ground state, convex hull, band structure, decoration, effective masses, density of states and phonons. Results obtained with ASR are automatically stored in a well-defined data format, which in addition to the results stores calculation metadata such as parameters and code versions. ASR has built-in support for workflows, however, MyQueue has been used for a more fine-grained control of the workflow.

The ASR chooses calculation parameters that should be well-converged by default. In general, these parameters depend on the particular quantity being calculated and will be provided later in their respective sections. However, for all calculations a plane-wave basis set with a cutoff energy of 800 eV was employed.

\section{Workflow}
An overview of the workflow can be seen in Figure~\ref{fig:workflow} and will be detailed chronologically in the following sections. In summary however, the workflow consists of two sub-workflows. First, the stability of the atomic structures is addressed and the resulting atomic configurations are classified in the "Structure and stability"-workflow. Subsequently, in the "Property"-workflow different physical properties of the atomic structures are calculated for later analysis.

\subsection{Structure preparation}

In the calculations for the isolated components, the atoms are arranged so that the periodic direction is along the $z$-axis. The unit cell projection on the $xy$-plane is rectangular with a side length chosen such that the distance between the atoms in neighboring cells are at least 16 \AA\ apart. During structure optimizations the unit cells are not relaxed in the $x$- and $y$-directions.

\subsection{Structure relaxation}
The atomic structures are determined by minimizing the total energy calculated with density-functional theory (DFT) with respect to the atomic coordinates and the unit cell vectors. The ground state calculations are performed with the PBE xc-functional \cite{Perdew:1996ug} and with the additional D3-correction \cite{Grimme:2010ij} to account for van der Waals interactions. A Monkhorst-Pack k-point grid with a sampling density of 6.0 \AA\ is employed and the Fermi-temperature for the smearing of the electronic occupation numbers is set to 0.05 eV. The atomic structures are relaxed so that the maximum force on the nuclei is below 0.01 eV/\AA\ and the stress components on the unit cell are below 0.002 eV/$\textrm{\AA}^3$. The energies of the isolated components are calculated with both PBE and PBE-D3, while the bulk structures are calculated with only PBE-D3, because we do not expect PBE to appropriately describe the interactions between the 1D components.

\begin{figure}
\centering
\includegraphics[width=0.8\linewidth]{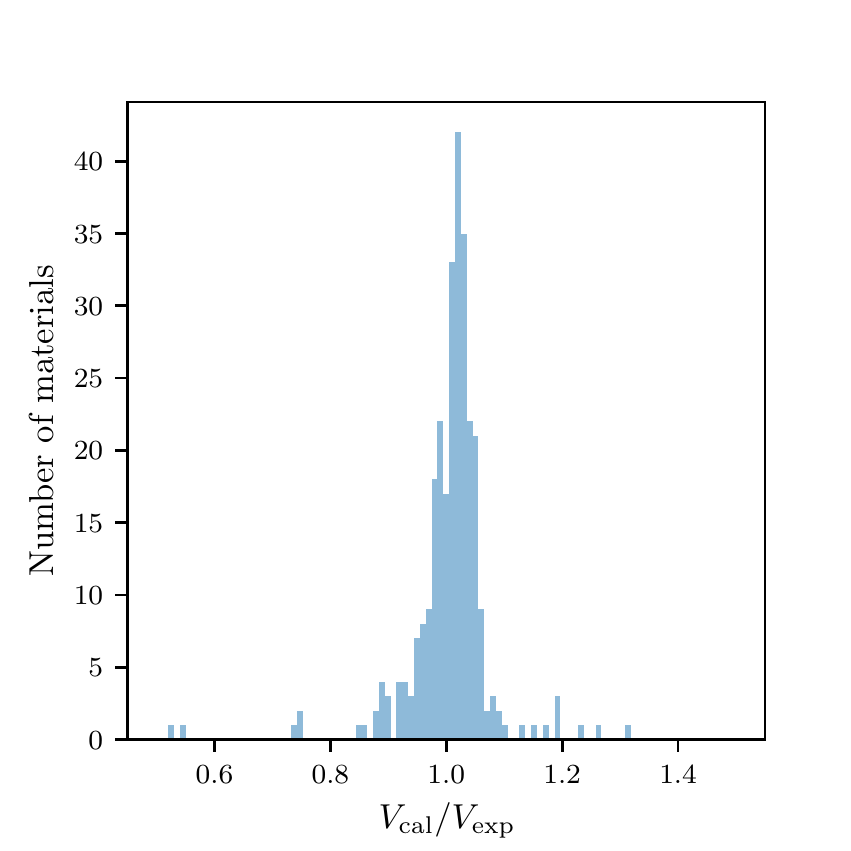}
\caption{The ratio between the volumes calculated with PBE-D3 and the experimental ones from ICSD and COD.}
\label{fig:volume}
\end{figure}

The 1D components in the bulk materials are bonded together by forces, which may be of van der Waals type. These are included in our calculations through the simple D3-correction. To investigate the performance of this approximation, we show in Fig.~\ref{fig:volume} the ratio between the calculated volumes after relaxation and the volumes from the initial configurations obtained from the databases ICSD and COD with experimentally determined lattice parameters. We see that the distribution includes a main peak and a number of outliers. Only considering the points in the interval $[0.9,1.1]$, we find an average value of 1.011 and a standard deviation of 0.035. The outliers are in some cases associated with considerable changes in structure. The two materials at the very low end of the histogram are \ch{Ag2S} and  \ch{CoO4S}. \ch{Ag2S} turns out to be a hypothetical structure, which was not removed in the initial selection steps of the database. The volume of the hypothetical structure is unphysically large, which explains the strong contraction during relaxation. The \ch{CoO4S} structure is reported with an experimental reference in COD, however, it seems that by mistake only half the atoms are actually included in the cif file. Other structures from the same reference contains twice the number of atoms. The volume of the more densely packed structure is in much better agreement with the calculated volume. The material at the high end of the histogram is \ch{ZrI3}. This material actually appears twice in the database. One of the structures is truly 1D with almost perfect agreement between the experimentally determined structure and the calculated one. The other structure actually lacks a reference article in COD and should have been removed in the initial selection. This structure exhibits an unrealistically small volume. There are a total of 6 materials in the database that occur twice.

\begin{figure}
\centering
\includegraphics[width=0.8\linewidth]{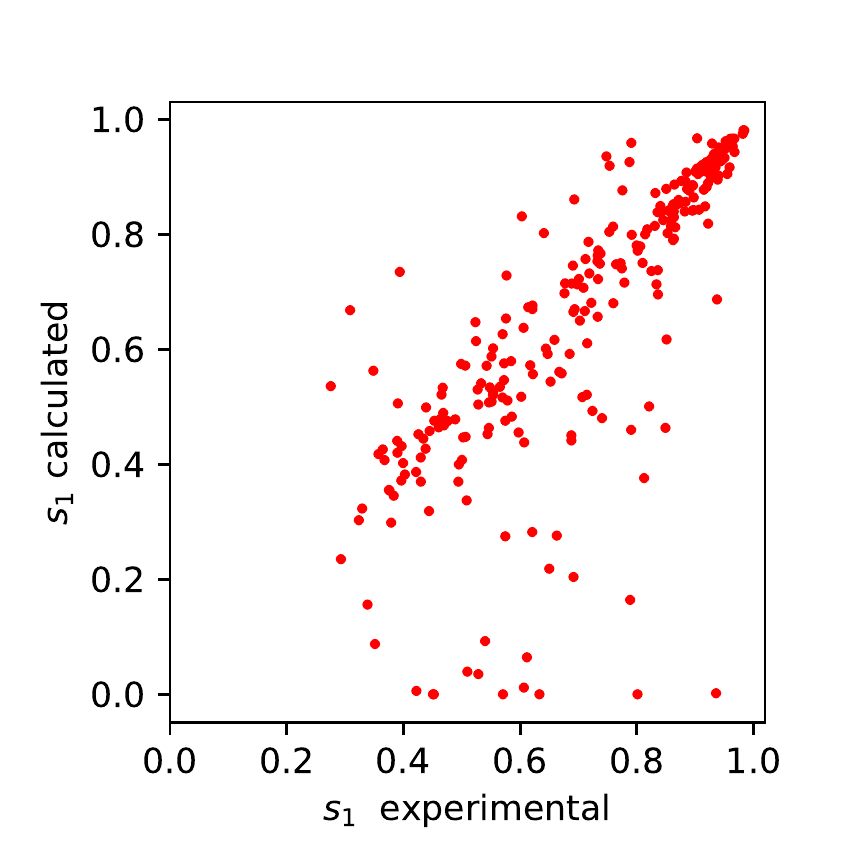}
\caption{ The scoring parameter calculated for the bulk materials in the core of the database relaxed with PBE-D3 versus the scoring parameter for the experimentally determined bulk structures.}
\label{fig:scoring}
\end{figure}

The structural changes caused by the relaxations can also be illustrated by the changes in the dimensionality scoring parameter $s_1$. Fig.~\ref{fig:scoring} shows a comparison between the scoring parameter calculated for the theoretically determined structures and the experimental ones. As expected the scoring parameters before and after relaxation are close although with a considerable spread and some significant outliers. As above the outliers are associated with major structural rearrangements. There are for example several materials including \ch{TlAlSiO4}, and \ch{NbBr5}, where the experimental $s_1$ is higher than 0.7 for which the calculated 1D scoring parameters almost vanish. As with the analysis of the volume changes, the main issue is bad entries in the experimental databases COD and ICSD. \ch{TlAlSiO4} appears in both ICSD (id: 89722) and COD (id: 9002425) based on the same experimental reference \cite{Kyono}, however, with two different sets of atomic coordinates. We expect the structure in ICSD with the $s_1$ very close to one to be wrong. The ICSD entry with id 35410 for \ch{NbBr5} exhibits also a value of $s_1$ close to one, but the computational relaxations result in a zero-dimensional molecular structure. However, the entry in ICSD has been superseded by another one (id: 67298), which does in fact have a high 0D score. 

There is furthermore a small group of materials with the calculated $s_1$ close to one and the similar experimental scoring parameter around 0.75. It turns out that these materials have significant values for the experimental scoring parameters involving zero-dimensional components, $s_0$ and $s_{01}$, so the values of $s_1+s_0+s_{01}$ are in fact close to unity.

\subsection{Magnetic classification}

\begin{table}
    \centering
    \caption{The magnetic
state for both the core and shell crystals structures. one bulk core and three bulk shell systems are missing because of convergence problems. While one 1D shell system is missing}\label{tab:magnetic_state}
    \begin{tabular}{ |l|r|r|r|r|}  
    \hline
Magnetic state     & Core 1D &Shell 1D&Core bulk & Shell bulk \\ \hline
Non-magnetic
&232 &462&241&471\\ \hline
Ferromagnetic
&54&65&45&56\\ \hline
Antiferromagnetic
&2&4&1&2\\ \hline
\end{tabular}
\end{table} 

Atomic structures are relaxed in both non-magnetic, ferromagnetic, and antiferromagnetic phases, and the phase with the lowest energy is stored in the database. Antiferromagnetic calculations are only carried out for systems with two metal atoms in the primitive unit cell. In Table~\ref{tab:magnetic_state} we show the resulting number of materials in the different magnetic phases. Clearly most of the materials are non-magnetic. However, it should be noted that only few material support antiferromagnetic ordering to begin with (ie. two metallic atoms in the primitive cell).

\subsection{Geometric classification}

The bulk materials can be classified by their symmetries and the corresponding space group as obtained with \emph{spglib} \cite{togo2018textttspglib}. This information is provided in the database.

The one-dimensional components can also be characterized by their line group symmetries, but here we shall discuss a more direct similarity measure between the different structures based on the root-mean-square-distance in coordinate space.

Consider two different structures with atomic coordinates $\vec{R}_i$, $i=1,2,\ldots N$ and $\vec{R}^\prime_i$, $i=1,2,\ldots N$. The root-mean-square distance (RMSD) is then defined as
\begin{equation}
    \textrm{RMSD} = \sqrt{\frac{1}{N} \sum_{i=1}^N |\vec{R}_i-\vec{R}^\prime_i|^2}.\label{eq:RMSD}
\end{equation}

In order to apply this expression to the one-dimensional components, several points have to be addressed. Firstly, we ignore the chemical identity of the atoms and consider only the coordinates. Secondly, two different 1D components may contain a different number of atoms, and in that case the systems are repeated along the $z$-direction so that both systems have the least common multiple of atoms. Thirdly, the two unit cells are scaled in the $z$-direction to get the same length. The coordinates in the perpendicular directions are not re-scaled. And finally, the distance has to be minimized with respect to translation and rotation of the two systems relative to each other, and the optimal mapping between the atoms in the two systems has to be identified. The translation and mapping components are adapted from the RMSD calculation method for bulk crystals described in Ref.~\onlinecite{Larsen:2017vj}, and the rotational alignment is found with a branch-and-bound algorithm.

\begin{figure}
\centering
\includegraphics[width=0.8\linewidth]{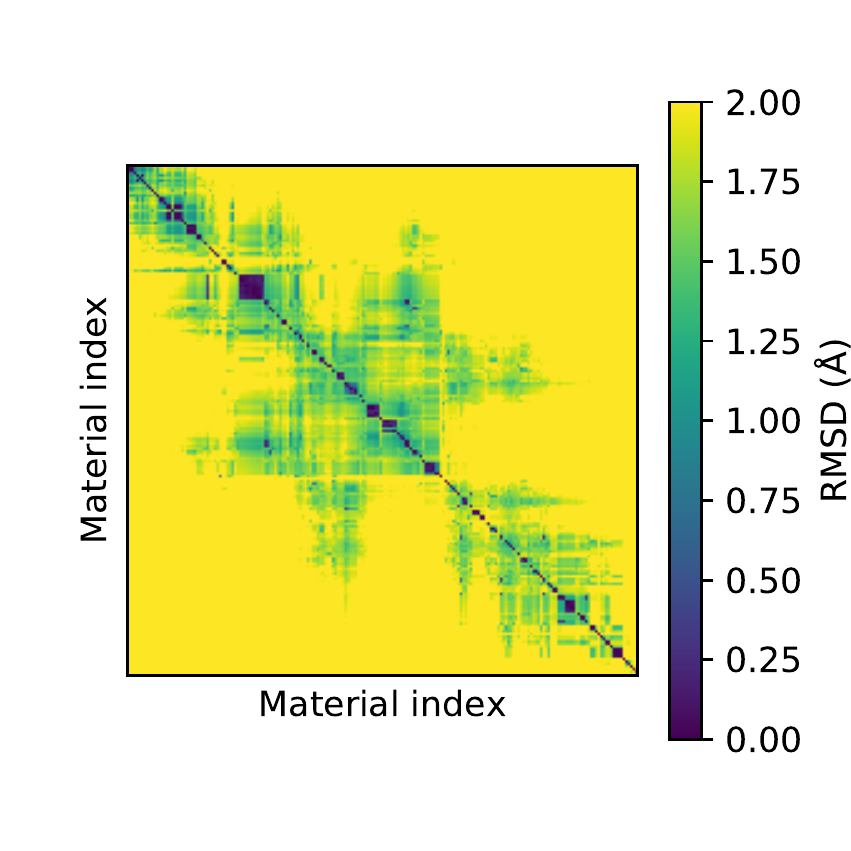}
\caption{The structure distance matrix of all materials having three or fewer different elements in the unit cell. The structures have been ordered using the rearrangement clustering method \cite{Zhang:2006wr} so that the dark squares on the diagonal reveal clusters of similar materials.}
\label{fig:RMSD}
\end{figure}

\begin{figure}
\centering
\includegraphics[width=0.8\linewidth]{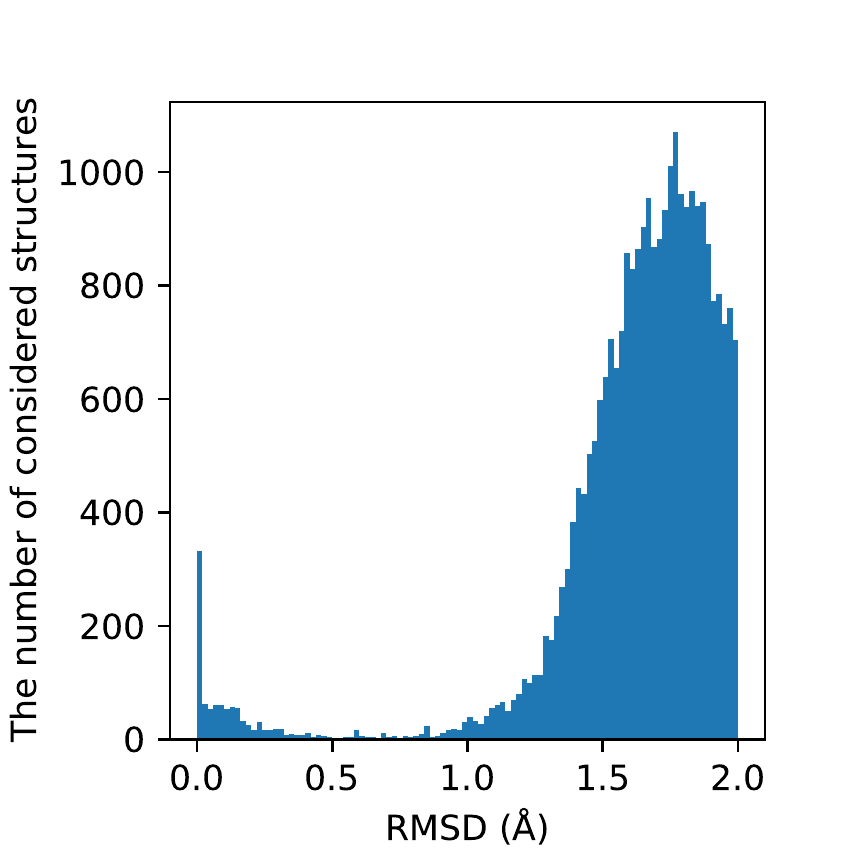}
\caption{Distribution of all distances in the distance matrix. The distances can clearly be separated in two groups with distance smaller or larger than 0.75 $\textrm{\AA}$.
}
\label{fig:histogram for dendrogramet}
\end{figure}

Figure~\ref{fig:RMSD} illustrates the RMSD distance matrix between all the 1D components in the core of the database. The materials have been optimally permuted using the rearrangement clustering method \cite{Zhang:2006wr}, so that clusters with similar structures appear dark along the diagonal.

We show in Figure~\ref{fig:histogram for dendrogramet} the distribution of distances from the distance matrix Figure~\ref{fig:RMSD}. The peak at zero distance corresponds to the diagonal in the distance matrix, and the height is just the number of structures considered. A clear double-peak distribution is seen, indicating that it makes sense to separate structures into some, which are close and others, which can be considered far away from each other. We take $R_0 = 0.75 \textrm{\AA}$ as a typical distance to separate the two groups.
\begin{figure}
\centering
\includegraphics[width=0.8\linewidth]{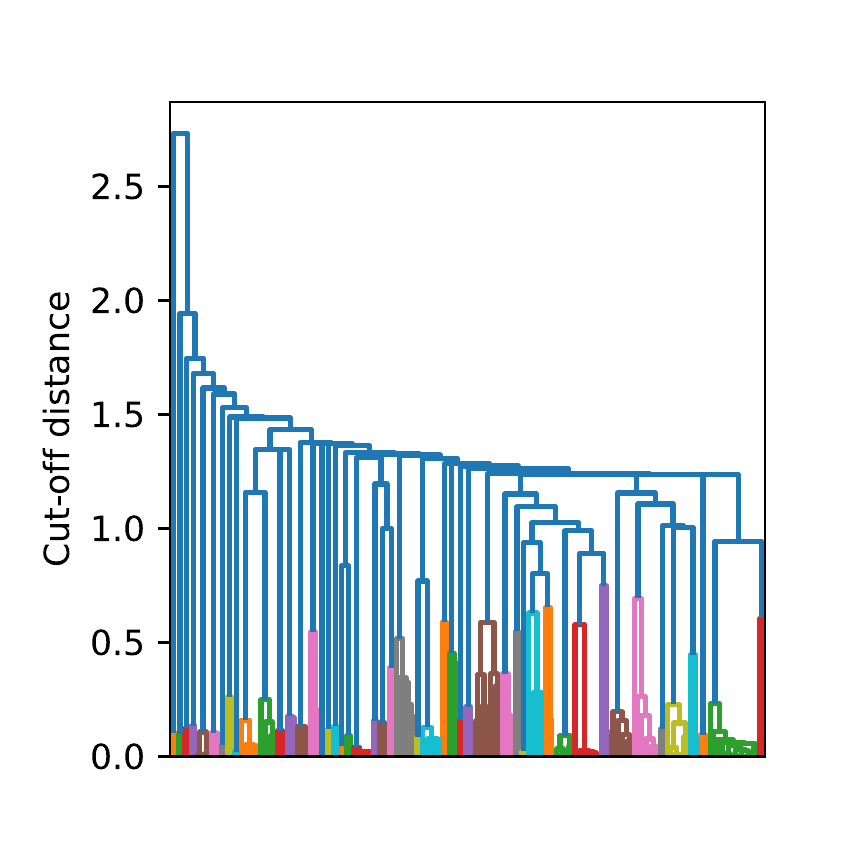}
\caption{A dendrogram of the clusters using single linkage. The clusters obtained by using a cutoff distance of 0.75 $\textrm{\AA}$ are shown in different colors.}
\label{fig:dendrogram}
\end{figure}
The grouping of the materials are further analyzed using single-linkage clustering. In this approach the clustering is a function of a cutoff distance. For a given value of the cutoff distance, two structures are considered related if their distance is smaller than the cutoff distance. This gives rise to an equivalence relation, where the equivalence classes constitute the clusters. The process can be visualized in a so-called dendrogram as shown in Figure~\ref{fig:dendrogram}. The cutoff distance is along the $y$-axis, and as the value is increased, smaller clusters join to form larger ones. For the value $R_0 = 0.75 \textrm{\AA}$ we find 48 groups. 27 of these contain three materials or more, and 13 groups have five materials or more. 
 \begin{table}
    \centering
    \begin{tabular}{|p{0.6\columnwidth}|p{\smallpic}|p{\smallpic}|}
    \hline
   Structure formula&  z direction&  y direction\\ \hline
  \emph{1}: \ch{ZrI3}, \ch{HfI3}, \ch{TiI3}, \ch{ZrBr3}, \ch{RuCl3}, \ch{TiCl3}, \ch{ZrCl3}, \ch{RuBr3}, \ch{ MoBr3}, \ch{ NbI3}, \ch{ Cs3O} &\parbox[c]{1em}{
      \includegraphics[width=\smallpic]{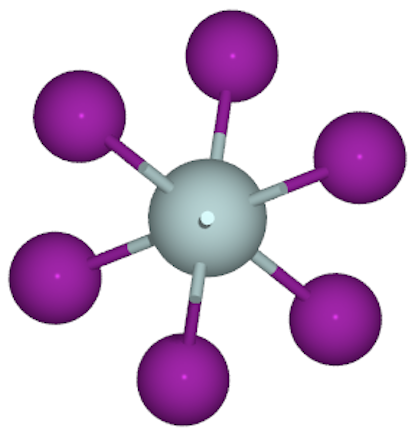}}&\parbox[c]{0.8em}{
      \includegraphics[width=\smallpic]{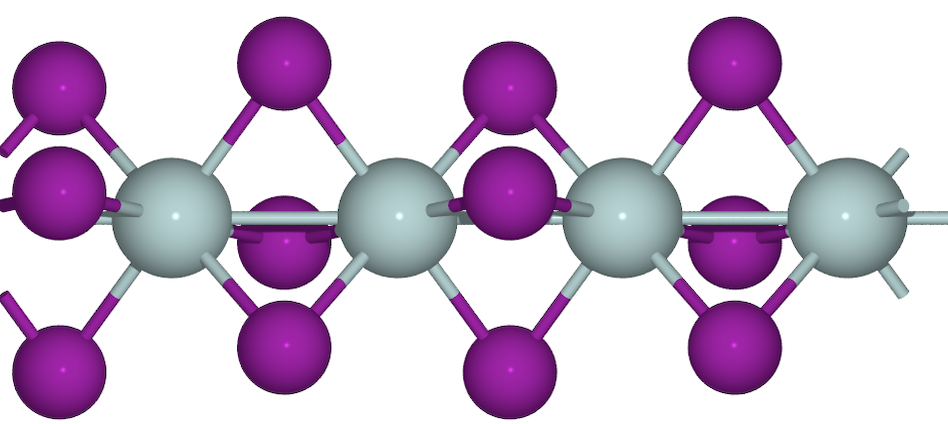}}
 \\ \hline
\emph{2}: \ch{    PdCl2}, \ch{  PdI2}, \ch{  CuCl2}, \ch{  CrBr2}, \ch{ CuBr2}, \ch{ CrI2}, \ch{ CrCl2}, \ch{PdBr2} &\parbox[c]{1em}{
      \includegraphics[width=\smallpic]{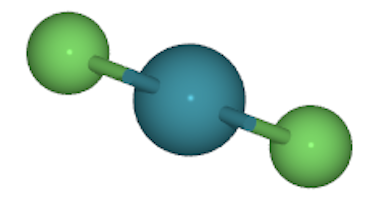}}&\parbox[c]{1em}{
      \includegraphics[width=\smallpic]{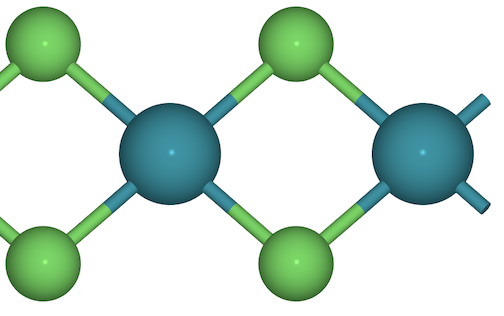}}\\ \hline
\emph{3}: \ch{VF5}, \ch{ CrF5}, \ch{ NbF10Sb}, \ch{ OsF4O}, \ch{  OsF3O2}, \ch{ OsF2O3}, \ch{ NbI5}, \ch{ ReF3O2}&\parbox[c]{1em}{
      \includegraphics[width=\smallpic]{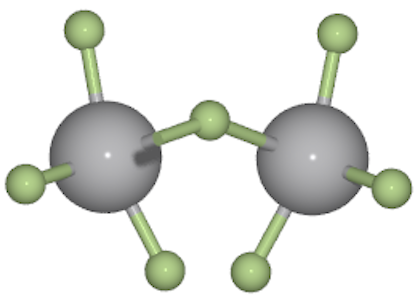}}&\parbox[c]{1em}{
      \includegraphics[width=\smallpic]{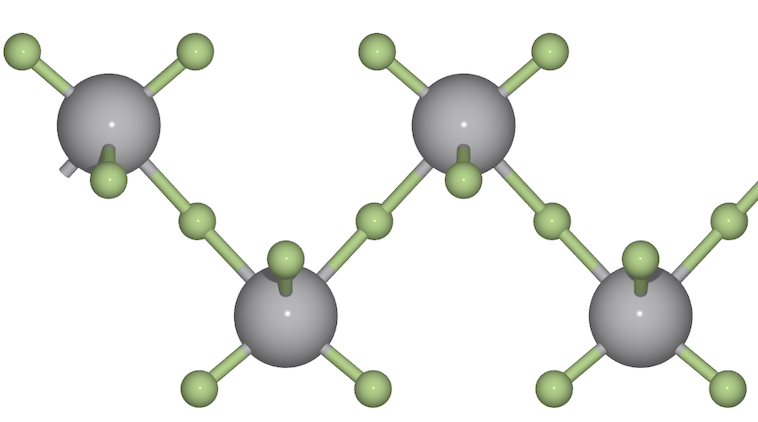}}
  \\ \hline
\emph{4}: \ch{  WBr4}, \ch{  WCl4}, \ch{  NbCl4}, \ch{  TaCl4}, \ch{  NbI4}, \ch{ TaCl2I2}, \ch{ OsCl4}, \ch{  PdTl2Se2} &\parbox[c]{1em}{
      \includegraphics[width=\smallpic]{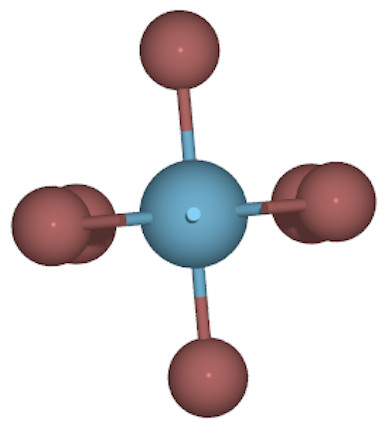}}&\parbox[c]{1em}{
      \includegraphics[width=\smallpic]{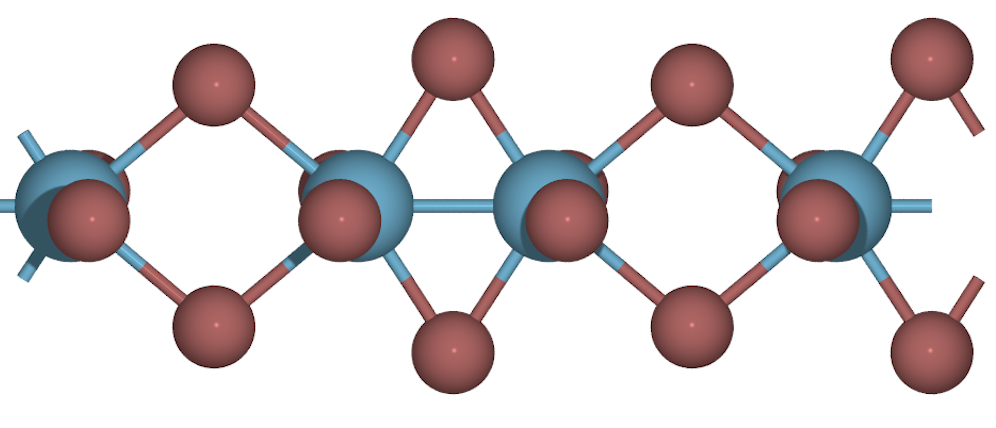}}
 \\ \hline
\emph{5}: \ch{    MoCl3S},
\ch{MoCl3Se},
 \ch{NbBr3Se},
\ch{NbI3Te},
\ch{ NbCl3Se},
\ch{ NbBr3Te},
\ch{ NbI3Se} &\parbox[c]{1em}{
      \includegraphics[width=\smallpic]{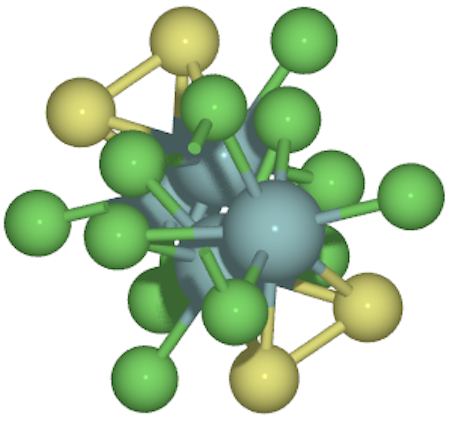}}&\parbox[c]{1em}{
      \includegraphics[width=\smallpic]{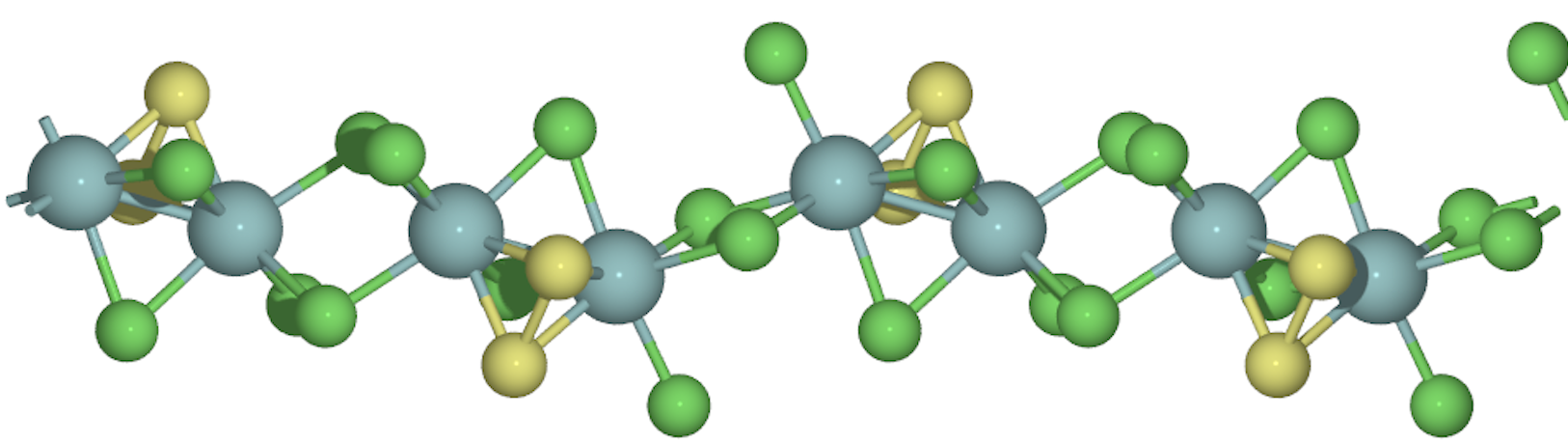}}
 \\ \hline
\emph{6}: \ch{  HfCl4}, \ch{  ZrCl4}, \ch{ ZrCl4}, \ch{TiI4}, \ch{ PtI4}, \ch{ OsBr4}, \ch{  Mo2Cl6O2} &\parbox[c]{1em}{
      \includegraphics[width=\smallpic]{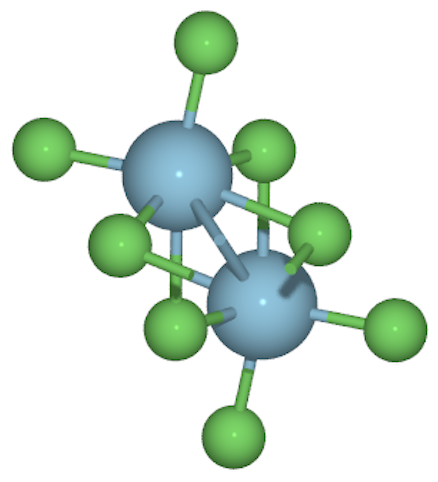}}&\parbox[c]{1em}{
      \includegraphics[width=\smallpic]{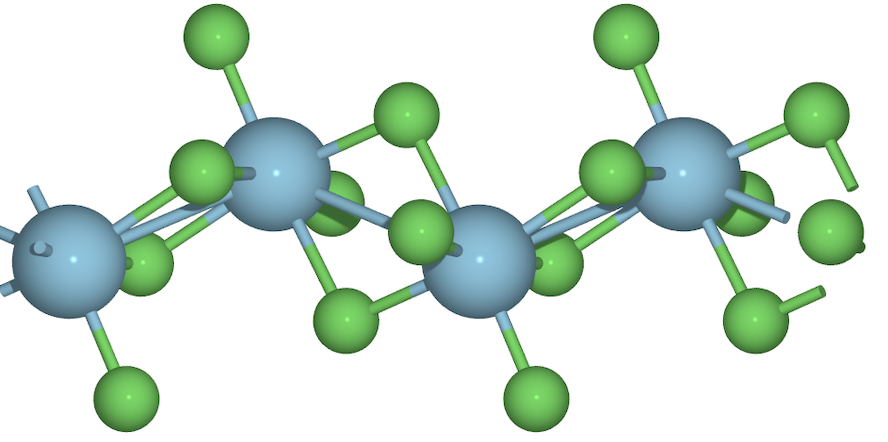}}
\\ \hline
\emph{7}: \ch{  Ta4SiTe4}, \ch{ FeTa4Te4}, \ch{FeNb4Te4}, \ch{  CoTa4Te4}, \ch{CrTa4Te4}, \ch{ NiTa4Te4}
 &\parbox[c]{1em}{
      \includegraphics[width=\smallpic]{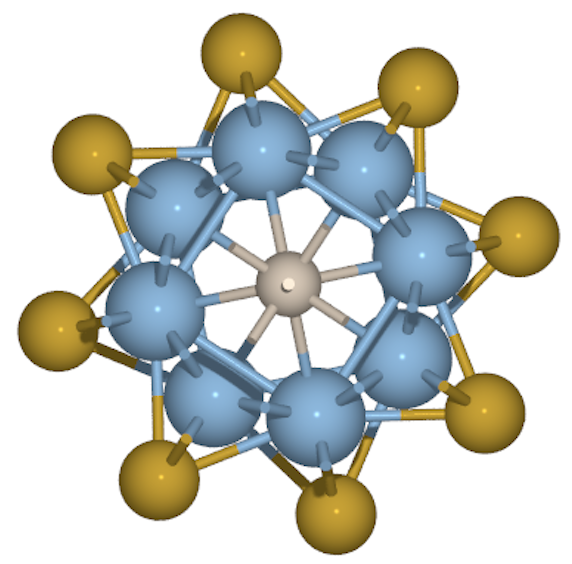}}&\parbox[c]{1em}{
      \includegraphics[width=\smallpic]{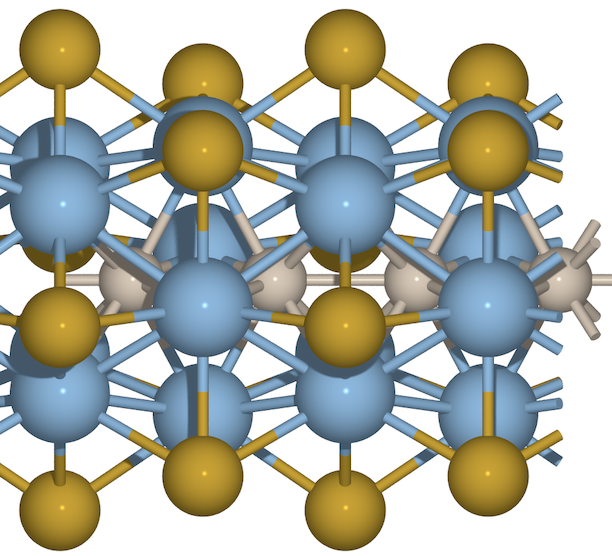}} \\ \hline
\emph{8}: \ch{BeCl2}, \ch{ Se2Si}, \ch{BeBr2}, \ch{BeI2}, \ch{ BPS4}, \ch{AlPS4} & 
 \parbox[c]{1em}{
      \includegraphics[width=\smallpic]{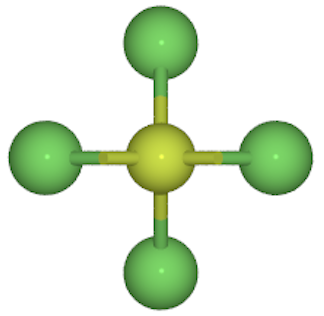}}&\parbox[c]{1em}{
      \includegraphics[width=\smallpic]{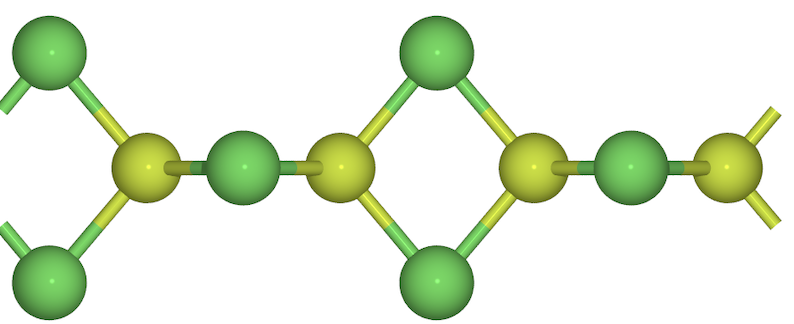}}\\ \hline
 \emph{9}: \ch{ BiIS}, \ch{ ISSb}, \ch{ BiBrS}, \ch{ BrSSb}, \ch{ BiBrSe}, \ch{ AsISe}  &\parbox[c]{1em}{
      \includegraphics[width=\smallpic]{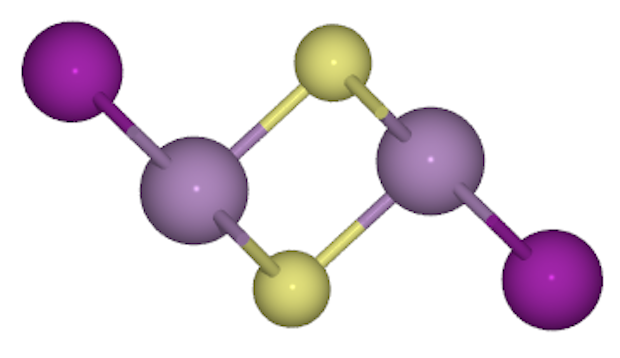}}&\parbox[c]{1em}{
      \includegraphics[width=\smallpic]{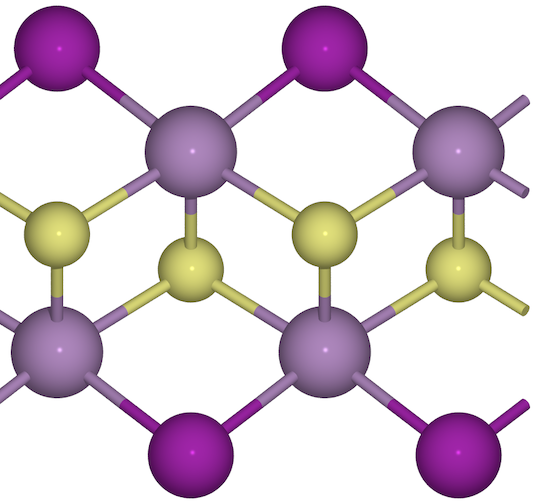}}
 \\ \hline
\emph{10}: \ch{   NbCl3O}, \ch{  NbBr3O}, \ch{  NbI3O}, \ch{  WCl3O}, \ch{  WI3O}, \ch{  CrF4}&\parbox[c]{1em}{
      \includegraphics[width=\smallpic]{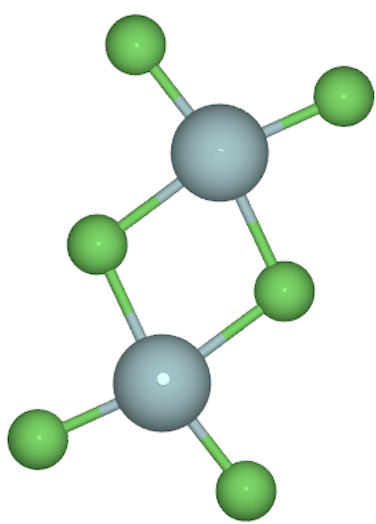}}&\parbox[c]{1em}{
      \includegraphics[width=\smallpic]{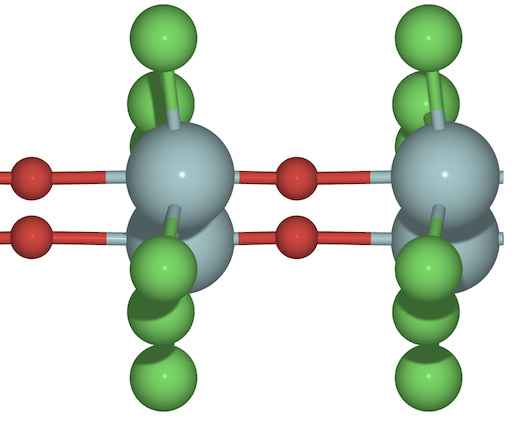}} \\ \hline
\emph{11}: \ch{ HfPbS3}, \ch{PbZrS3}, \ch{ HfSnS3}, \ch{ SnZrS3}, \ch{ Sn2S3}, \ch{  As2Te3} &\parbox[c]{1em}{
      \includegraphics[width=\smallpic]{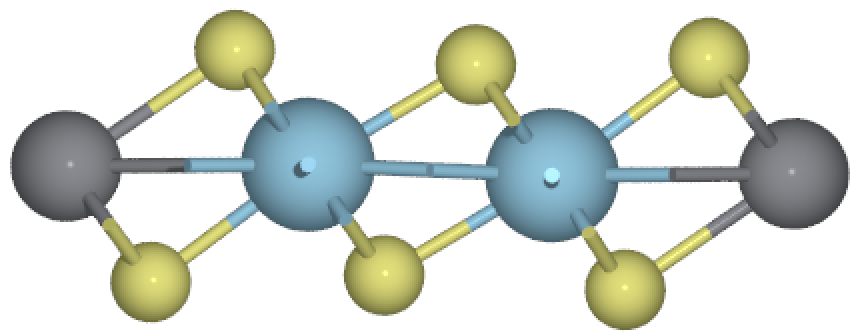}}&\parbox[c]{1em}{
      \includegraphics[width=\smallpic]{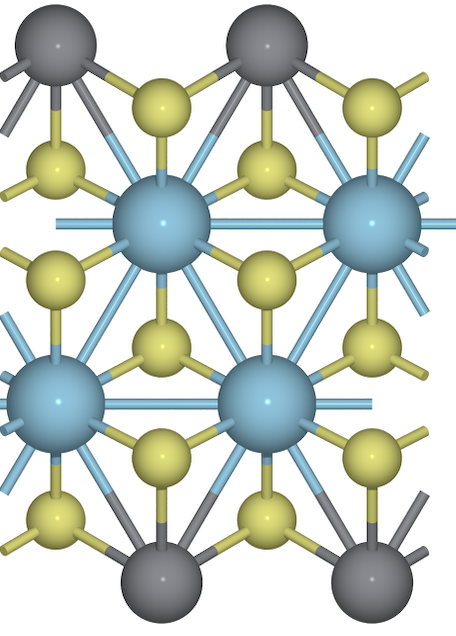}}
  \\ \hline
\emph{12}: \ch{    AgCN}, \ch{   AuCN}, \ch{   AgI}, \ch{   HgO}, \ch{SnS} &\parbox[c]{1em}{
      \includegraphics[width=\smallpic]{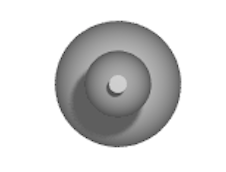}}&\parbox[c]{1em}{
      \includegraphics[width=\smallpic]{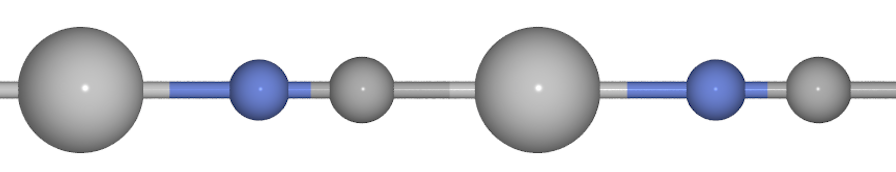}}
 \\ \hline
\emph{13}: \ch{     CrS3Sb}, \ch{  CrSbSe3}, \ch{ TiGeS3}, \ch{PbSnS3}, \ch{  InS3Sb }& \parbox[c]{1em}{
      \includegraphics[width=\smallpic]{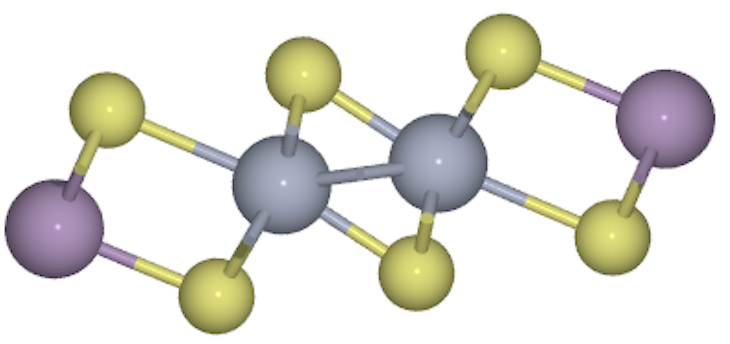}}&\parbox[c]{1em}{
      \includegraphics[width=\smallpic]{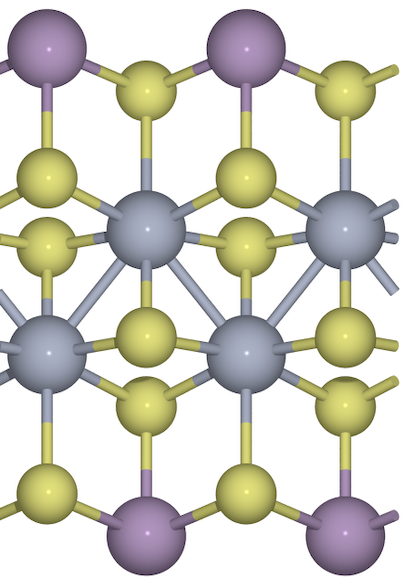}}
 \\ \hline

\end{tabular}
\caption{Crystal structure clusters found using the dendrogram. Clusters with more than five materials are shown in the table.}\label{tab:structure_type}
\end{table}
We show the 13 groups with five or more materials in Table~\ref{tab:structure_type}. In some cases the grouping is clearly related to the groups in the periodic table. For example, the largest group with 11 materials contains \ch{Ti2X6} with \ch{X} = \ch{Cl}, \ch{Br}, and \ch{I}. However, it is not from the outset clear that \ch{Cs3O} would belong to the same group. The components \ch{ZrI3}, \ch{RuBr3}, \ch{RuCl3}, and \ch{TiI3} actually appear twice in the database because they are extracted from different bulk structures. For example, the two bulk structures of \ch{ZrI3} (with COD numbers 5910009 and 74648) are made up of the same one-dimensional components, but they are stacked differently in the two bulk materials.
It is interesting to compare our structural groups with the ones discussed recently by Zhu et al. \cite{zhu_spectrum_2021, Cheon:2017kn}. They identify seven groups with four or more materials using the Structure Matcher module of Pymatgen. Three of their groups with representative materials \ch{HfI3}, \ch{BiSI} and \ch{HfPbS3} correspond to our groups number 1, 9 and 11. Their groups with \ch{CoH4(ClO)2} and with \ch{ZnH4(CO3)2} we also find, but with less than five materials. In their group with \ch{Ti(AlBr4)2} the three other systems have more than 20 atoms in the unit cells and are therefore not considered here. Finally, Zhu et al. finds a group with \ch{InGaTe2} and four other materials. Three of these we find to have low values for $s_1$, and the two other ones have warnings in ICSD about high-pressure structures. We therefore do not consider these five materials here.

\subsection{Thermodynamic stability and convex hull}
A key quantity signifying the stability of a compound is the heat of formation $\Delta H$. For a binary compound with composition \ch{A_nB_m} the heat of formation per atom is defined as $\Delta H = (E(\ch{A_nB_m})-n E(\ch{A}) -m E(\ch{B}))/(n+m)$, where $E(\ch{A_nB_m})$ is the total energy per unit cell of the compound, and $E(\ch{A})$ and $E(\ch{B})$ are the energies per atom in the standard states for the elements \ch{A} and \ch{B}. This is easily generalized to compounds with more elements. If the heat of formation is positive, the compound can be expected to disintegrate into the constituent elements, while if it is negative it will be stable at low temperatures. This principle can be generalized to decomposition of a compound into other (not necessarily elemental) materials and can be investigated through the so-called convex hull construction. For a given set of reference materials, $M_i$, $i = 1,2, \ldots N$, with heats of formation $H(M_i)$ and a given composition, for example \ch{A_nB_mC_p}, the mix of materials with that composition and the lowest heat of formation will be positioned at the convex hull. We shall denote the energy of a material relative to the convex hull as $\Delta H_\textrm{hull}$.

As reference materials we use all relevant elemental, binary and ternary materials at the convex hull in the OQMD database \cite{Kirklin:2015cr}, where we have recalculated the structures and energies with GPAW using the PBE-D3 functional with the same parameter settings as the for the other energy calculations in this work. The set of reference materials is further supplemented with the bulk materials from both the core and shell of the database. The calculations of all the reference materials are also included in the online version of the database.

\begin{figure}
\centering
\includegraphics[width=0.8\linewidth]{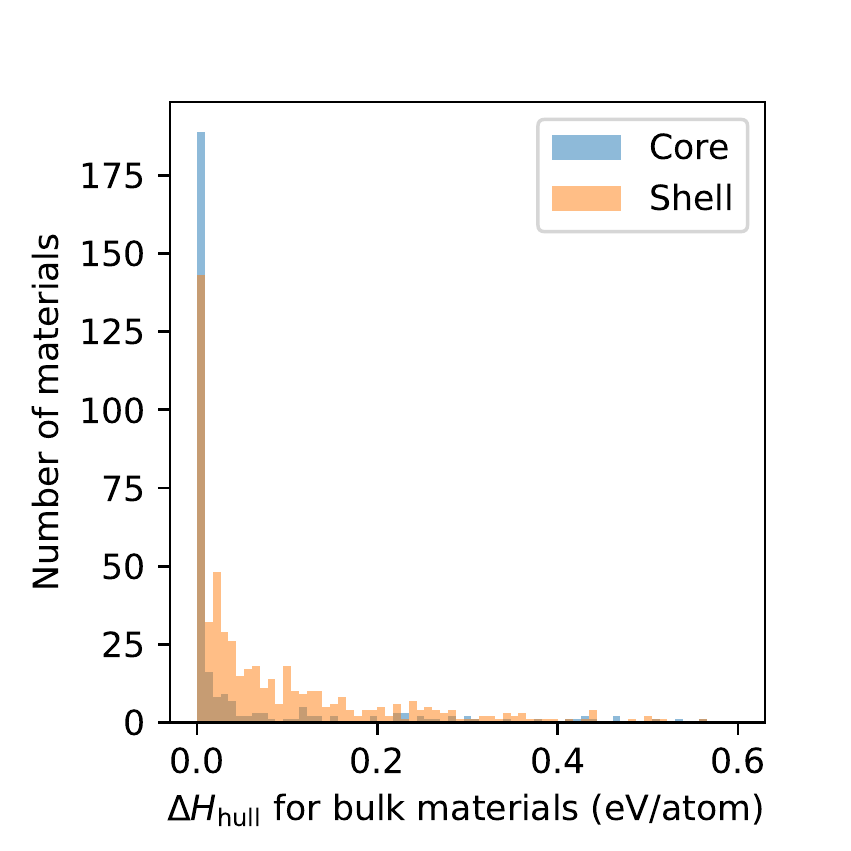}
\caption{Distribution of the energies above the convex hull, $\Delta H_\textrm{hull}$ calculated with PBE-D3 for the bulk materials in the core and the shell of the database.}
\label{fig:Ehull}
\end{figure}

\subsubsection{Bulk materials}
We first look at the bulk materials. Fig.~\ref{fig:Ehull} shows the distribution of calculated energy differences, $\Delta H_\textrm{hull}$ between the materials and the convex hull. Since the materials themselves are part of the reference systems for the hull construction all energies are positive. We see that the materials in the core of the database have low energies, in most cases below 0.1 eV. This is to be expected since the core materials are experimentally synthesized materials extracted from ICSD and COD, so they are known to be (meta-)stable. The distribution for the shell materials is somewhat wider, but with most of the materials with energies below 0.2 eV/atom. It can therefore be expected that a considerable number of these materials can in fact be synthesized in their bulk form.

\begin{figure}
\centering
\includegraphics[width=0.8\linewidth]{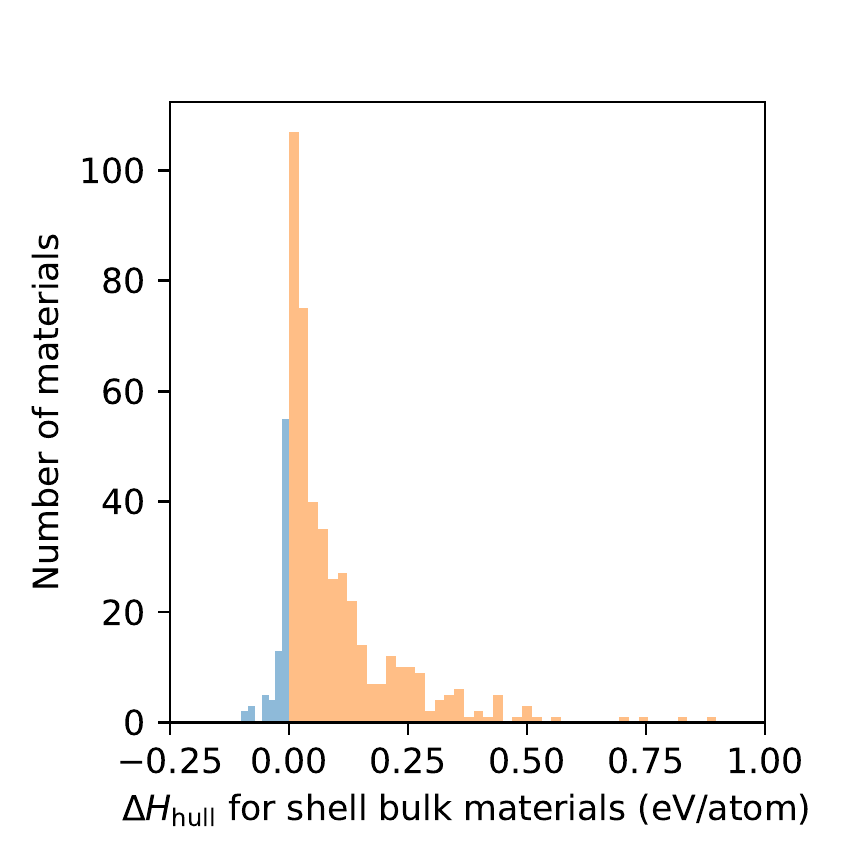}
\caption{Distribution of above-the-hull energies for the bulk materials in the shell as calculated with PBE-D3. Here, the set of reference materials for the convex hull construction consists of the unary and binary materials from OQMD and the materials in the core of the database.}
\label{fig:shell_ehull}
\end{figure}

To analyze the stability of the materials in the shell of the database further, Fig.~\ref{fig:shell_ehull} shows the distribution of above-the-hull energies, but now relative to a reference database consisting of only the materials from OQMD and from the core of the database. As can be seen a substantial fraction of these (82 materials) exhibit negative energies indicating that these could be genuinely new materials not being investigated before. A manual check shows that 55 of these materials do in fact not appear in ICSD or COD. These materials are listed in Table~\ref{tab:stable_shell} together with the calculated scoring parameter values for $s_1$. Most of the materials maintain the one-dimensional character as indicated by the large values of $s_1$. The table also includes the electronic energy gaps as calculated with PBE. As shell materials they are all derived by element substitution from core materials. The parent material including its structure can be found in the online version of the database.

\begin{table}
    \centering
   \begin{tabular}{ |l|r|r|l|r|r|}  
    \hline
    Name&$s_1$& Gap &Name&$s_1$& Gap \\
 \hline
\ch{Al3ClSe3} & 0.94 & 1.9 &\ch{GeTiSe3}&0.71&0.1 \\
 \hline
\ch{W2Br8N} & 0.93 & 0.6& \ch{Cl2Se3} & 0.70 & 1.6 \\ \hline
\ch{Al3BrSe3} & 0.92 & 2.0& \ch{Br2Se3}&0.68&   1.33 \\
 \hline
\ch{MoBr3S2} & 0.92 & 1.1 &\ch{PdTl2S2} & 0.62 & 0.0 \\
 \hline
\ch{MoCl3Se2} & 0.92 & 1.1&\ch{PdTl2S2}         &0.62& 0.0\\\hline

\ch{Ga3ClSe3} & 0.91 & 1.8 &\ch{AgNb2F12} &0.61     &0.82\\
 \hline

\ch{ Al3ISe3 } &      0.9   &   1.9&\ch{SnZrSe3} & 0.58 & 0.7 \\ \hline

\ch{MoBr3Se2} & 0.9 & 1.0 &\ch{PdTl2Te2} & 0.57&  0.00\\ \hline
\ch{Mo2Br8N} & 0.9 & 0.0&\ch{SnBrI} & 0.55 & 2.4 \\
 \hline
\ch{W2Br7N} & 0.9 & 0.0 &\ch{InSbSe3} & 0.52 & 0.9\\
 \hline
 \ch{Mo2Br7N   } &     0.9 &   0.0& \ch{Hg2O5S} & 0.50 & 1.1\\
 \hline
 \ch{Ga3BrSe3} & 0.89 & 1.8 &\ch{PbZrSe}&0.49&0.78\\
 \hline
\ch{In3BrTe3} & 0.89 & 1.2 &\ch{PdBr2Se} & 0.44 & 0.9\\
 \hline
\ch{W2I8N} & 0.88 & 0.0 &\ch{BrSe2 }& 0.43&    0.70 \\
 \hline
 \ch{Al3BrTe3} & 0.87 & 1.8&\ch{RhBr3Se6} & 0.42 & 1.6\\ \hline
 \ch{Al3ClTe3} & 0.87 & 1.7 &\ch{IrBr3Se6} & 0.41 & 1.4\\ \hline
\ch{In3ClTe3} & 0.87 & 1.1 &\ch{RhI3Se6} & 0.40 & 1.0\\ \hline
\ch{Ga3BrTe3} & 0.86 & 1.2&\ch{IrI3Se6} & 0.38 & 0.8 \\ \hline
\ch{FeTa4Se4} & 0.86 & 0.0 &\ch{RhCl3Se6} & 0.33 & 1.8 \\
 \hline
\ch{TaF10Sb} & 0.85 & 4.9&\ch{IrCl3Se6} & 0.33 & 1.7\\
 \hline
 \ch{Ga3ClTe3} & 0.85 & 1.2  &\ch{BrSbTe} & 0.31 & 0.9\\ \hline
\ch{Li2O5Se2} & 0.84 & 3.5&\ch{LiPSe2}&0.24&0.0 \\
 \hline
\ch{W2I7N} & 0.84 & 0.0& \ch{AuBr4P} & 0.22 & 1.7\\
 \hline
 \ch{PdCl2S  } &    0.82 &   1.44&\ch{WBr3N} & 0.14 & 1.3\\
 \hline
 \ch{PdBr2S} & 0.82 & 1.1&\ch{Na2O5Se2} & 0.06 & 3.2 \\
 \hline 
 \ch{Mo2I8N } & 0.81 &   0.0&\ch{PdBr2Te} & 0.0 & 0.6 \\
 \hline
 \ch{ IPSe  } &0.80  &  1.5&\ch{FeO4Te} & 0.0 & 0.0 \\
 \hline
 \ch{WI3N} & 0.74 & 0.6&&& \\
 \hline
\end{tabular}
\caption{A list of the bulk materials from the shell with energies below the convex hull. The materials are new in the sense that they are not present in ICSD and COD. The listed electronic energy gaps are calculated with PBE.}\label{tab:stable_shell}
\end{table}

\subsubsection{One-dimensional components}
We now turn to the isolated 1D components in the database. These are again divided in core and shell systems with the core systems being obtained by extraction of 1D components from the bulk materials in the core, and the shell systems by element substitution in the core systems. 

We evaluate the low-temperature stability of an isolated 1D component by comparing its energy, $E_\textrm{1D}$, to the energy of the parent bulk material, $ E_\textrm{bulk}$. This is done using PBE-D3 to take van der Waals bonding in the parent material into account. We shall term this energy difference per atom the \emph{separation} energy $\Delta E_\textrm{sep} = E_\textrm{1D}/N_\textrm{1D}-E_\textrm{bulk}/N_\textrm{bulk}$, where $N_\textrm{1D}$ and $N_\textrm{bulk}$ are the number of atoms in the unit cells for the isolated component and the bulk system, respectively.

\begin{figure}
\centering
\includegraphics[width=0.8\linewidth]{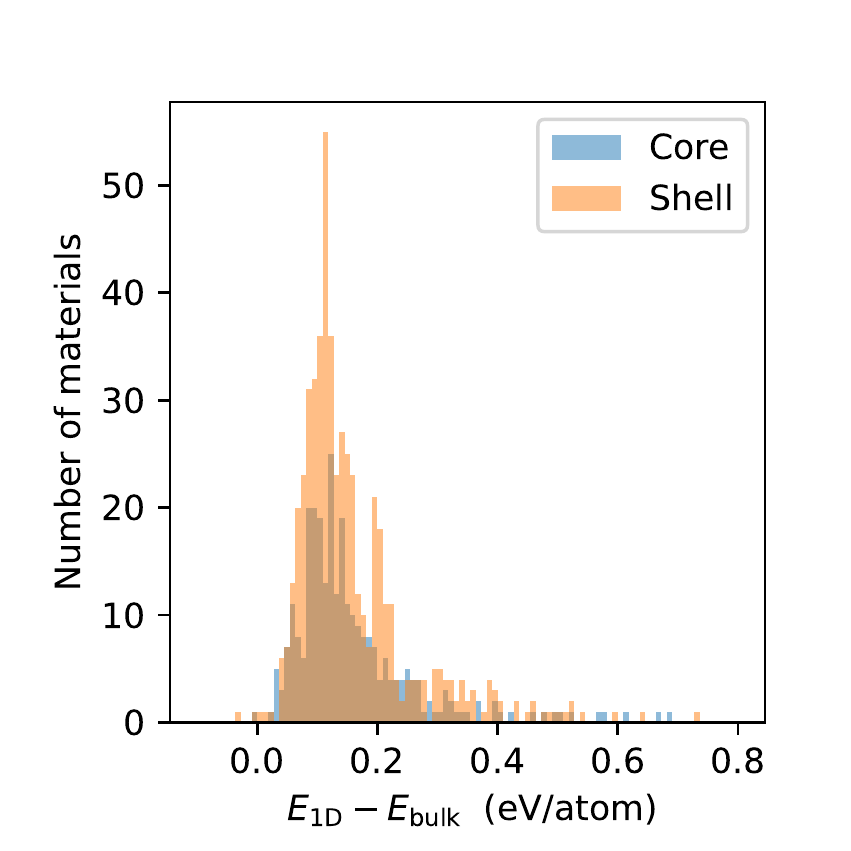}                \caption{The distribution of separation energies defined as the difference between the relaxed 3D bulk material and the 1D system per atom calculated using the PBE-D3 functional.}               \label{fig: 1D and 3D}
\end{figure} 

Figure~\ref{fig: 1D and 3D} shows a histogram of the separation energies for both the core and the shell materials of the database. Typically the separation energy is only about 0.1 eV per atom indicating that the bonds, which are broken, are fairly weak as expected from the  selection procedure of the materials based on the dimensionality scoring parameter. Interestingly, the separation energies of the shell materials do not seem to be higher than for the core materials, providing the possibility of identifying new stable one-dimensional components.

Two-dimensional layers may in some cases be produced by exfoliation from the corresponding bulk systems, and potentially one-dimensional materials could be made in a similar fashion \cite{muratov2020synthesis,kim2018mechanical,lipatov2018quasi}. In the case of the two-dimensional systems, it has been suggested that the degree to which the exfoliation process is possible can be estimated by the so-called exfoliation energy, $E_\textrm{xf}$ \cite{Mounet:2018ks}. The exfoliation energy is identical to the above defined separation energy, but instead of being evaluated per atom, it is normalized to the area of the two-dimensional component. Mounet et al. \cite{Mounet:2018ks} have suggested a threshold value of 35 $\textrm{meV}/\textrm{\AA}^2$ for the exfoliation energy to identify easily exfoliable two-dimensional materials. We note that the scoring parameter can also in itself be regarded as a measure of exfoliability, and it has been shown that experimentally exfoliated two-dimensional materials exhibit high values of $s_2$ \cite{Larsen:2019cf}.

\begin{figure}
\centering
\includegraphics[width=0.8\linewidth]{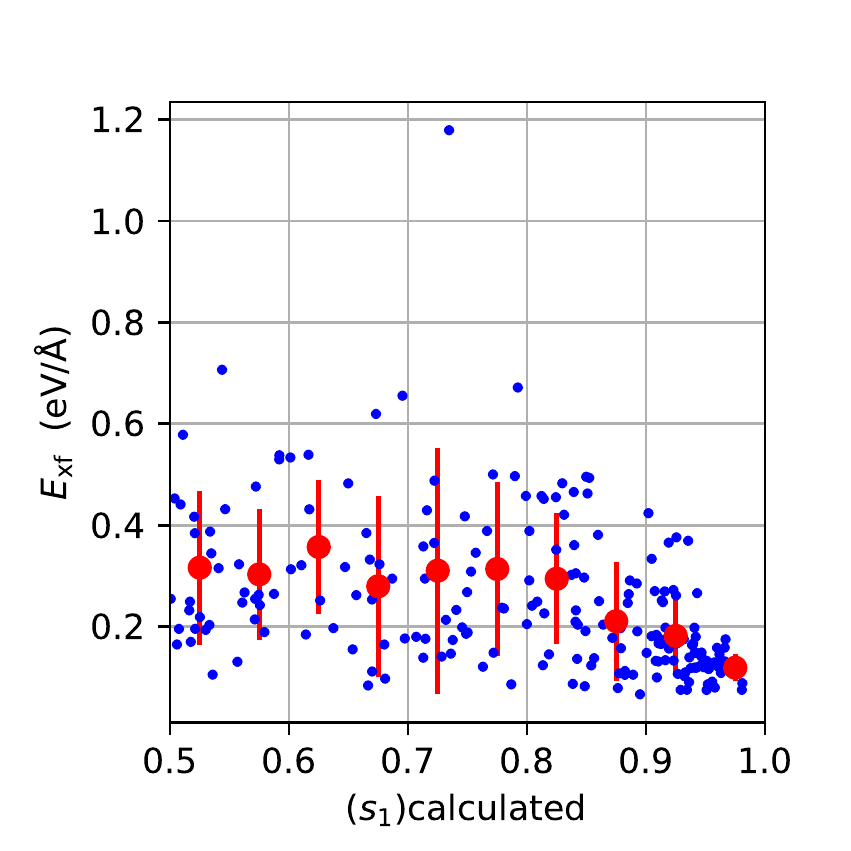}
\caption{The figure shows calculated exfoliation energies $E_\textrm{xf}$ versus the calculated scoring parameter $s_1$ for all materials in the core of the database. The red points and vertical bars show the average values and standard deviations for points in the intervals [0.5, 0.55], [0.55, 0.60] etc. for $s_1$. The two different measures of exfoliability are clearly correlated for large values of $s_1$.
}
\label{fig:s1_cal}
\end{figure}

The concept of exfoliation energy is easily generalized to one dimension, where the natural normalization is now per length of the one-dimensional component. It is thus related to the separation energy per atom as $E_{xf} = E_{sep} N_\textrm{1D} /L$, where $N_\textrm{1D}$ is the number of atoms in the unit cell, and $L$ is the length of the one-dimensional component for one unit cell. An alternative measure for how easy the exfoliation process might be for one-dimensional materials is the scoring parameter $s_1$, and in Figure~\ref{fig:s1_cal} we show a comparison of the two. They are clearly correlated in particular for large values of $s_1$. 

\begin{figure}
\centering
\includegraphics[width=0.8\linewidth]{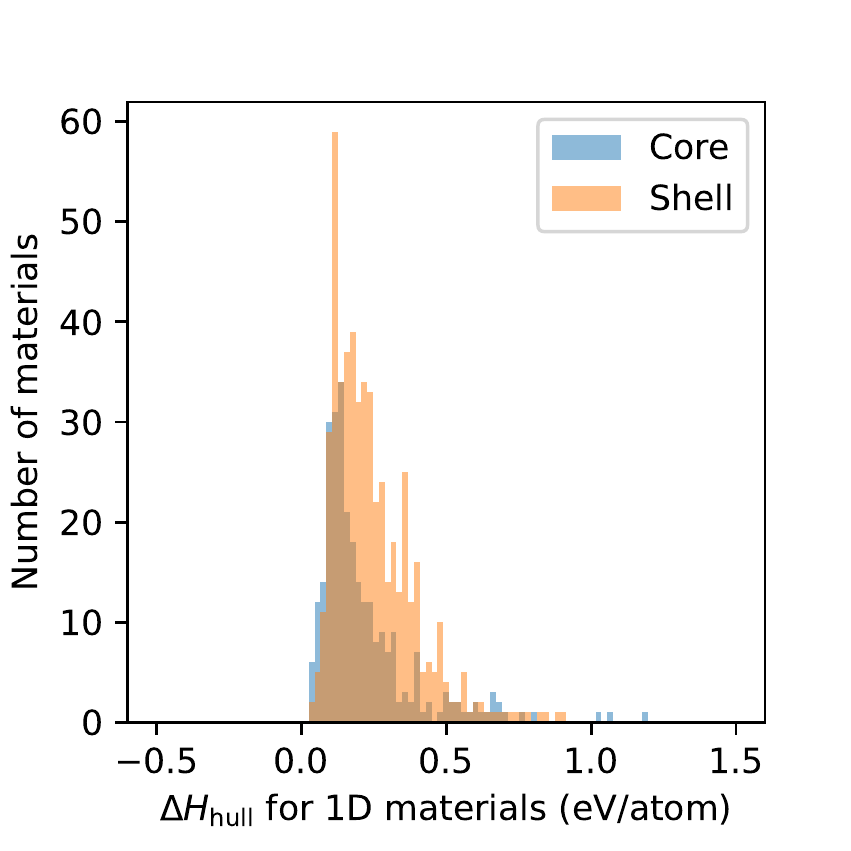}                \caption{The energy above convex hull distribution for the extracted 1D materials.}               \label{fig:Ehull_1D}
\end{figure}

The stability of a one-dimensional component can also be directly compared to the convex hull as $\Delta H_\textrm{hull,1D} = E_\textrm{sep} + \Delta H_\textrm{hull,bulk}$, where the last term is the energy above the hull for the corresponding 3D system. A histogram of this quantity is shown for all components in Figure~\ref{fig:Ehull_1D}. The shell components are on the average ~0.1-0.2 eV less stable than the core components. This shift is mainly due to the difference in the stability of the bulk materials (Figure~\ref{fig:Ehull}), while the separation energies are similarly distributed for the bulk and shell systems (Figure~\ref{fig: 1D and 3D}). 

\begin{figure}
\centering
\includegraphics[width=0.8\linewidth]{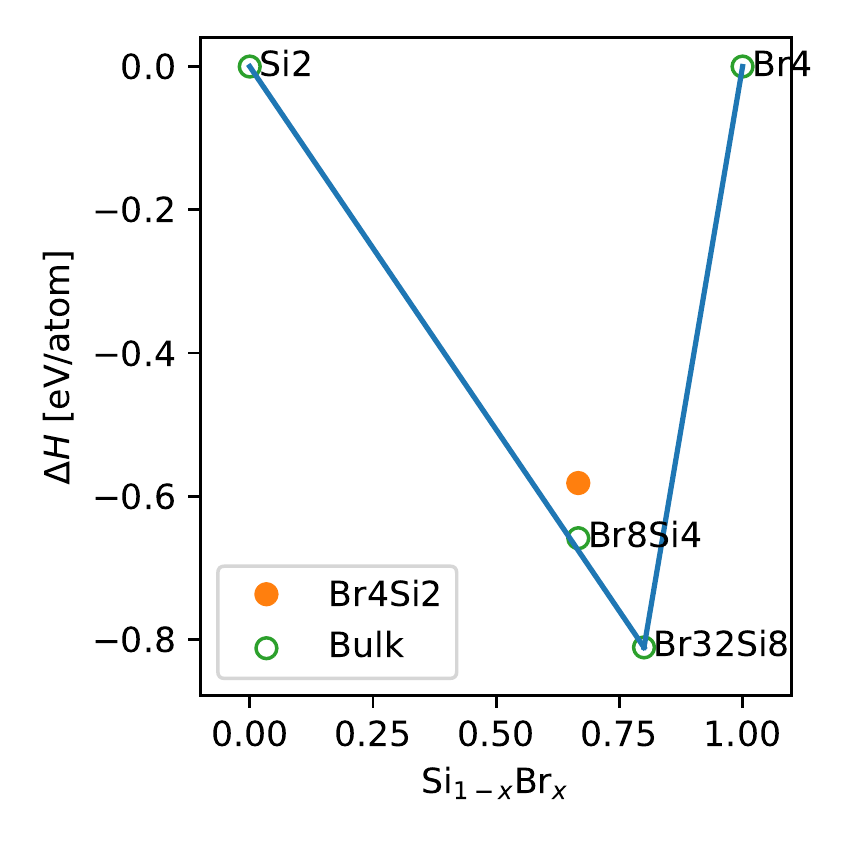}
\caption{A convex hull plot of a 1D material compound, with the bulk reference structures in green, and the 1D component in orange.}
\label{fig:hull}
\end{figure}

For all one-dimensional components the stability relative to the corresponding bulk material and the reference systems for the hull can be visualized in a hull plot as shown in Figure~\ref{fig:hull} in the case of $\ch{Br2Si}$. The figure illustrates both the energies relative to the hull, and the separation energy between the one-dimensional component and the corresponding bulk system. Hull plots are available for all investigated systems in the online database.

Considering the accuracy of the calculations and the possibility of meta-stability, we label the components with energies less than 0.2 eV/atom above the convex hull with the thermodynamic stability ``high''. This is similar to the classification used in C2DB for two-dimensional materials \cite{Haastrup:2018ca}.

\subsection{Phonons}

We calculate the $\Gamma$-phonon frequencies for the isolated components using the finite displacement method in a unit cell which is doubled along the direction of the 1D structure. The calculations are performed by constructing the force constant matrix using finite displacements of 0.01 \AA\ of the atomic positions. The ground state calculation parameters used here are identical to the parameters used during relaxation.
The phonon calculations are mostly used to investigate whether imaginary frequencies (corresponding to negative eigenvalues for the dynamical matrix) appear. Imaginary frequencies are an indication that the structure is not dynamically stable and will deform into another more stable configuration. Care must be taken to separate possible imaginary modes from the three zero-modes corresponding to translation of the system as a whole. In the database we label a material to be dynamically unstable if the Hessian has eigenvalues smaller than -0.01 meV/Å$^2$. If we focus on the components, which we label to be of high thermodynamical stability (i.e. components having a total energy less than 0.2 eV/atom above the convex hull) we find 183 components in the core of the database and 242 in the shell. Out of these 149 and 172 are labeled as also dynamically stable in the core and shell, respectively.

\subsection{Property workflow}

\subsubsection{Ground state}
Except for the calculation of the density of states, subsequent steps in the property workflow is based on a ground state calculation employing a k-point sampling density of 12 Å and the PBE exchange-correlation functional. This density ensures that reciprocal space local quantities like band extrema are precisely determined. Furthermore, an accurate ground state reduces the error-accumulation of the subsequent workflow steps that are based on this ground state calculation.

\subsubsection{Work function and band extrema}
We calculate the work function and band extrema when they are well-defined. The work function is defined as the difference between the Fermi level and the vacuum level, and as such is only defined for metals. Similarly, we define the conduction band minimum (CBM) and valence band maximum (VBM) as relative to the vacuum level.

Additionally, we also calculate the "direct" CBM and VBM. These are defined by locating the positions in k-space that minimizes the difference between the conduction band and valence band energies. Similarly, these are also measured with respect to the vacuum level.

In practice, we approximate the vacuum level of an isolated 1D component by the value of the Kohn-Sham potential in the space between the one-dimensional systems as far away from the atoms as possible. It is not possible to extract the vacuum level from bulk systems. These quantities are extracted from the ground state calculation that was the first step of the property workflow.

\subsubsection{Band gaps}
The energy band gaps are calculated (for non-metallic systems) as the energy difference between the CBM and the VBM. The direct band gap is calculated as the difference between the direct CBM and VBM. These are calculated both for the 1D and bulk systems.

\begin{figure}
\centering
\includegraphics[width=0.85\linewidth]{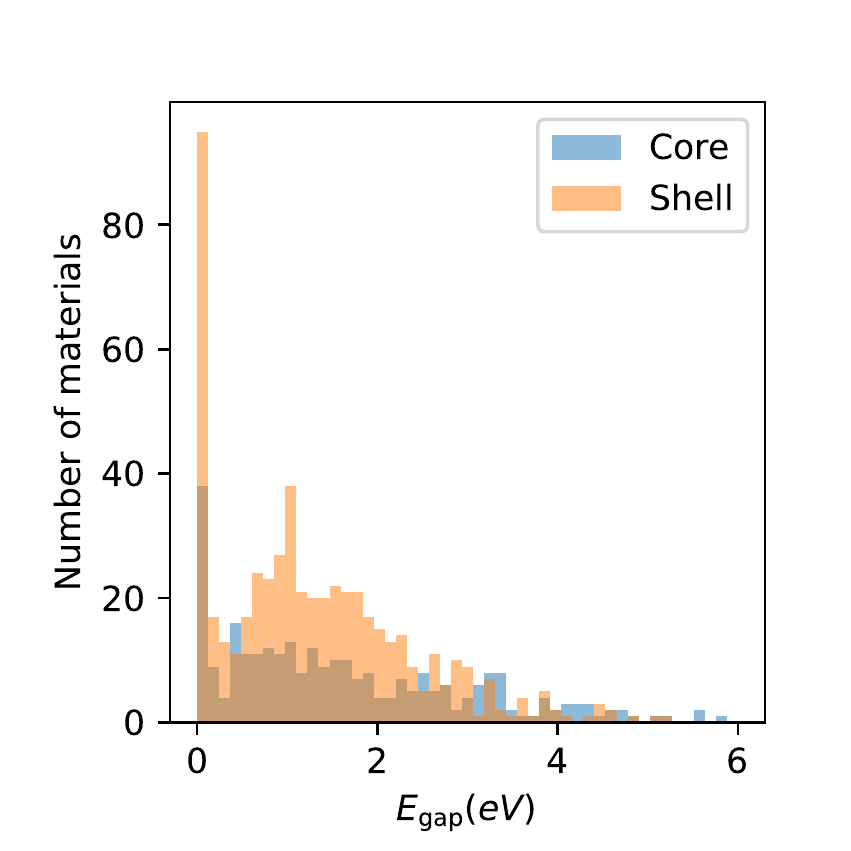}
\caption{The distribution of energy gaps for all the core and shell 1D materials as calculated with PBE.}
\label{fig:PBE_BS}
\end{figure}

Figure~\ref{fig:PBE_BS} shows the distribution of band gaps for the one-dimensional components as calculated with PBE. The narrow peak close to zero shows the number of metallic systems. The majority of the systems are clearly semiconducting and many have band gaps in the visible or ultraviolet range. The distribution for core and shell systems are fairly similar.

\begin{figure}
\centering
\includegraphics[width=1\linewidth]{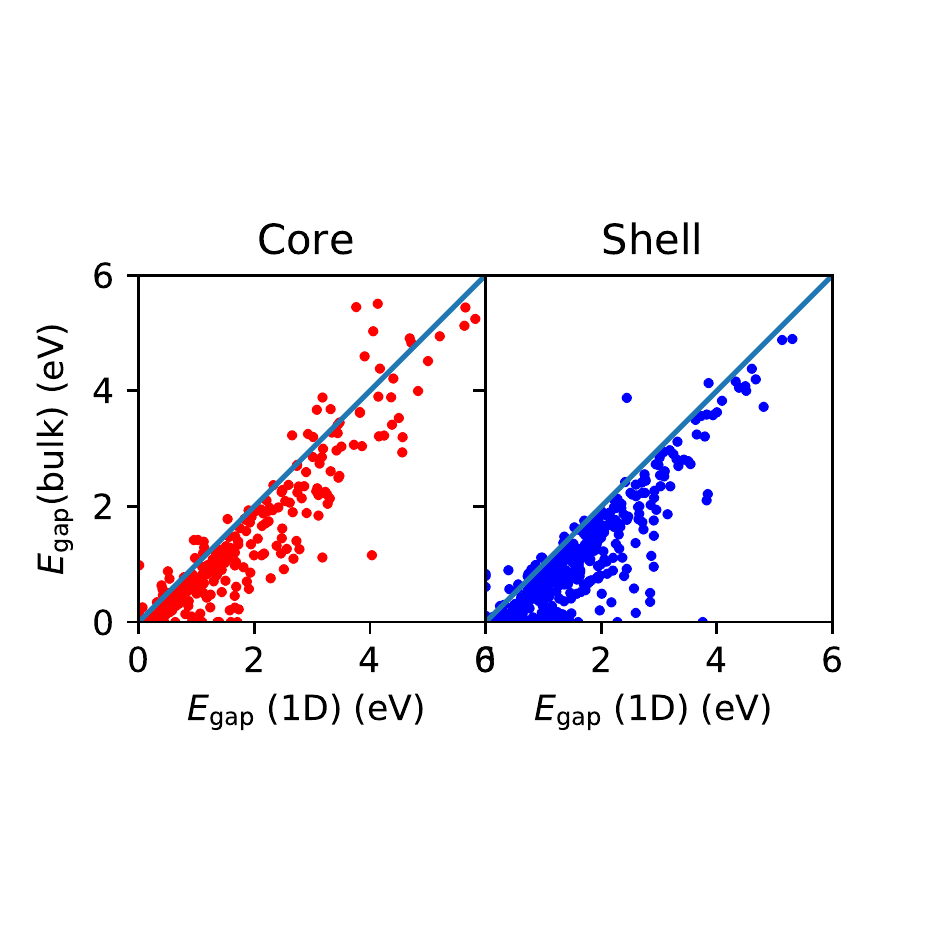}
\caption{Comparision between the PBE calculated band gaps for the bulk systems and the one-dimensional components.}
\label{fig:gaps_1D_3D}
\end{figure}

It is well known for 2D materials that both the size and the character of the band gap (\emph{i.e.} whether it is direct or indirect) may be different in single layers compared to bulk systems. In Figure~\ref{fig:gaps_1D_3D} we show a direct comparison of the calculated band gaps for the isolated components versus the gaps in the corresponding bulk materials. As might be expected the additional confinement present in the isolated components gives rise to an increase of the band gap for many systems.

\subsubsection{Band structure}
The one-dimensional band structures of the bulk materials and the 1D components are calculated from the Kohn-Sham energy levels using PBE and non-selfconsistent HSE06 \cite{Heyd:2003eg}. The electronic density obtained in the ground state calculation is kept fixed and a path in k-space is sampled with 400 points. For the bulk systems we use the paths defined by Setyawan and Curtarolo \cite{setyawan2010high} and for the one-dimensional system we consider the line between the $\Gamma$-point and the zone boundary. The band structures calculated for \ch{ISSb}  in both the bulk structure and as a one-dimensional component are shown in Figure~\ref{fig:band_structures}.

\subsubsection{Electronic density of states}
The electronic density of states is calculated using the tetrahedron interpolation method which in one dimension reduces to the trapezoidal rule for integration. This is particularly important for one-dimensional systems, which exhibits strong van Hove singularities at critical points diverging as the inverse square root of the energy for a parabolic band, i.e., $1 / \sqrt{E - E_C}$ where $E_C$ is the critical point energy. A simple point summation would converge poorly. In practice a dense k-point sampling is still required even when employing the trapezoidal rule, and the current study has used a density of 50 \AA$^{-1}$.

\begin{figure}
     \centering
         \includegraphics[width=0.475\columnwidth]{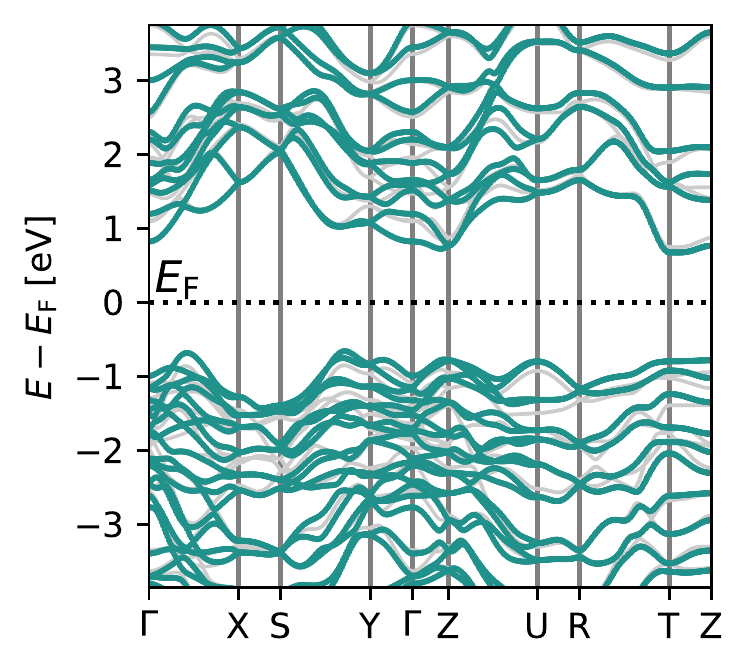}
         \includegraphics[width=0.475\columnwidth]{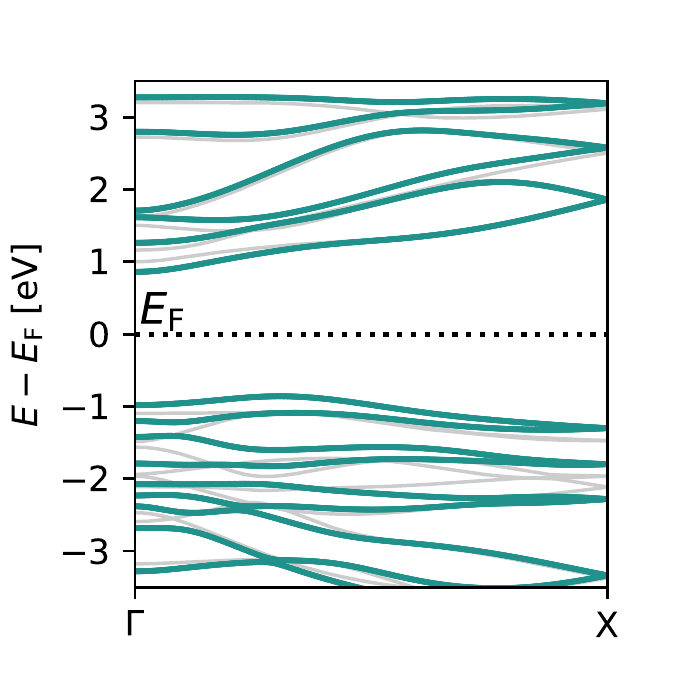}
         \includegraphics[width=0.475\columnwidth]{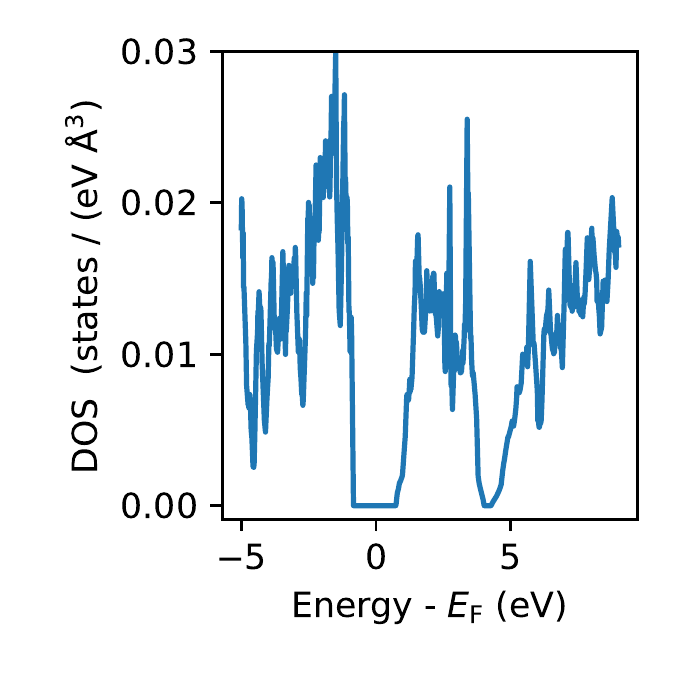}
         \includegraphics[width=0.475\columnwidth]{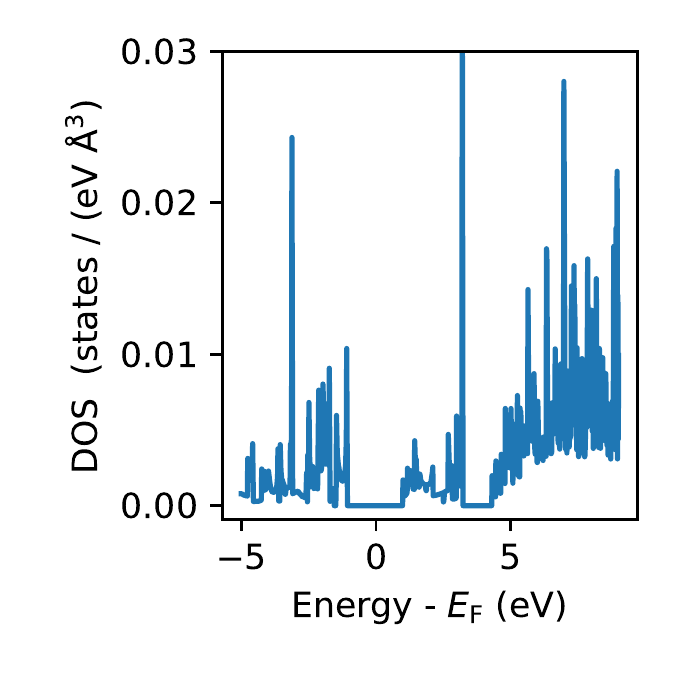}
        \caption{Band structure and electronic density of states for \ch{ISSb} for  both the bulk system (left figures) and the one-dimensional components (right figures). \ch{ISSb} belongs to group 9 in Table~\ref{tab:structure_type}.}
        \label{fig:band_structures}
\end{figure}

The resulting bulk and 1D densities of states for \ch{ISSb} are shown in Figure~\ref{fig:band_structures}. The van Hove singularities clearly appear as pronouned peaks in the density of states of the 1D system.

\subsubsection{Effective masses}
The effective masses of the one-dimensional components have been calculated by fitting parabolas to the band structure close to the VBM and CBM. One challenge here is that several bands may be present and potentially also cross each other close to the extrema. We identify the different bands by describing each electronic state at a given k-point and energy by a ``fingerprint'' consisting of the projections of the states onto the local PAW projectors. In principle, the fingerprint would vary continuously along a band, and the electronic states at different $k$-points are therefore joined into bands so that neighboring states have fingerprints, which are as close as possible. 

\begin{figure}
     \centering
         \includegraphics[width=0.9\columnwidth]{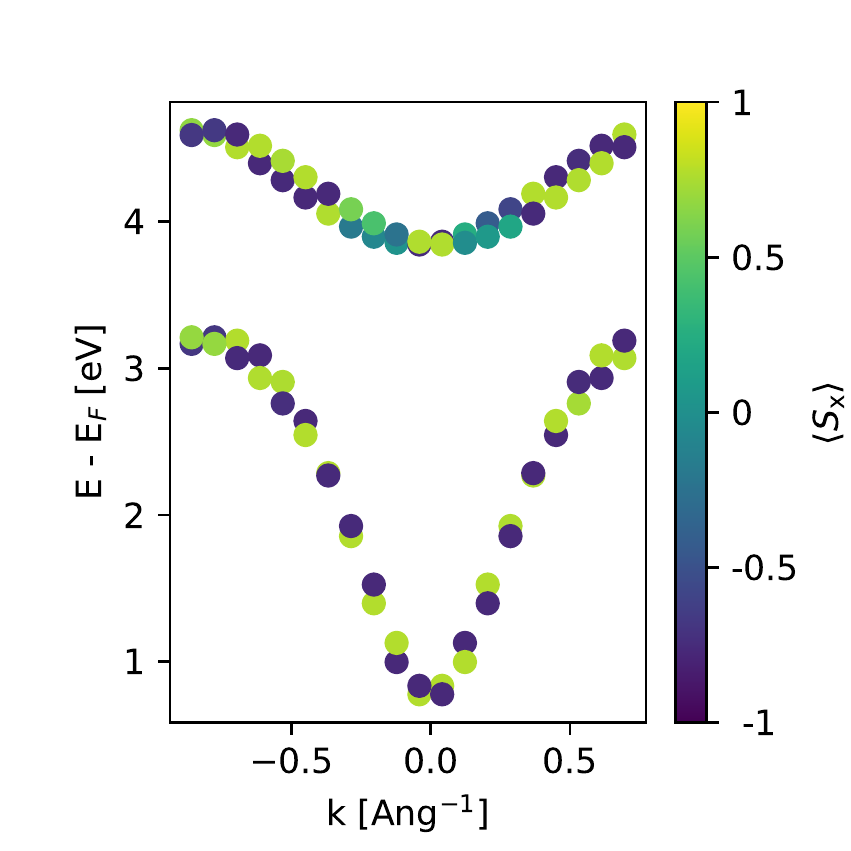}
         \includegraphics[width=0.9\columnwidth]{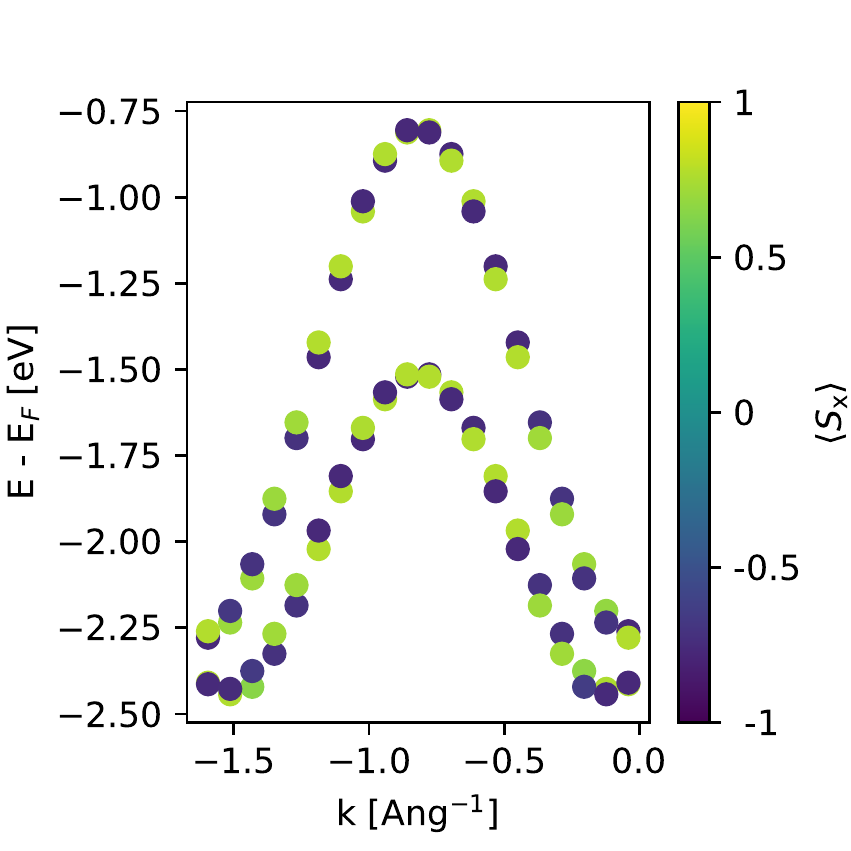}
        \caption{A zoom of the band structure of the one-dimensional component of \ch{HgO}. The upper figure shows the conduction band and the lower figure the valence band. The magnitude of the x-component of the spin is represented by the colors of the circles. Small Rashba splittings are observed in the bands.}
        \label{fig:masses}
\end{figure}

An example of a mass determination including the parabolic fits can be seen in Figure~\ref{fig:masses}.

\section{Majorana bound states}
In particle physics, a Majorana fermion is a fermion that is its own antiparticle\cite{majorana1937teoria}. In condensed matter physics, a Majorana bound state (MBS) is an unusual kind of quasiparticle, i.e. an excited state, whose creation operator equals its annihilation operator ($\hat \gamma^\dagger  =\hat \gamma$)\cite{majorana}. Such quasiparticles can emerge in the form of zero-energy bound states occurring at topological defects, e.g. interfaces, domain boundaries, or vortices of a topological superconductor \cite{fu2008superconducting, beenakker2013search, alicea2012new, flensberg_engineered_2021}. The latter are typically characterized by broken time-reversal symmetry and $p$-wave pairing for the superconducting state. Due to their intrinsically nonlocal nature (see below) and non-abelian exchange statistics, MBS are considered as candidates for low-decoherence quantum information processing such as fault-tolerant quantum computing\cite{kitaev2003fault,nayak2008non}.  

One of the simplest models in which MBSs appear is the so-called Kitaev model\cite{kitaev2001unpaired}, which is a tight-binding representation of a 1D $p$-wave superconductor. Since $p$-wave superconducting pairing couples electrons of the same spin (in contrast to $s$-wave pairing, which couples opposite spins), the Kitaev model considers spinless electrons. It can be shown that the Kitaev model hosts two MBSs located at each end of the chain. The physical content of a MBS is not intuitive as it is represented by a Hermitian creation/annihilation operator; thus e.g. the notion of occupation does not apply to a MBS. However, by superposing the two Majorana end states one obtains a fermionic state, $\hat f=(\hat \gamma_1+i\hat \gamma_2)/2$. The ground state of the Kitaev model is doubly degenerate, corresponding to the fermionic state being either empty or occupied. The key property of this fermionic state is that it is protected against decoherence processes due to the spatial separation of its MBS constituents.   

\begin{figure}
     \centering
         \includegraphics[width=0.98\columnwidth]{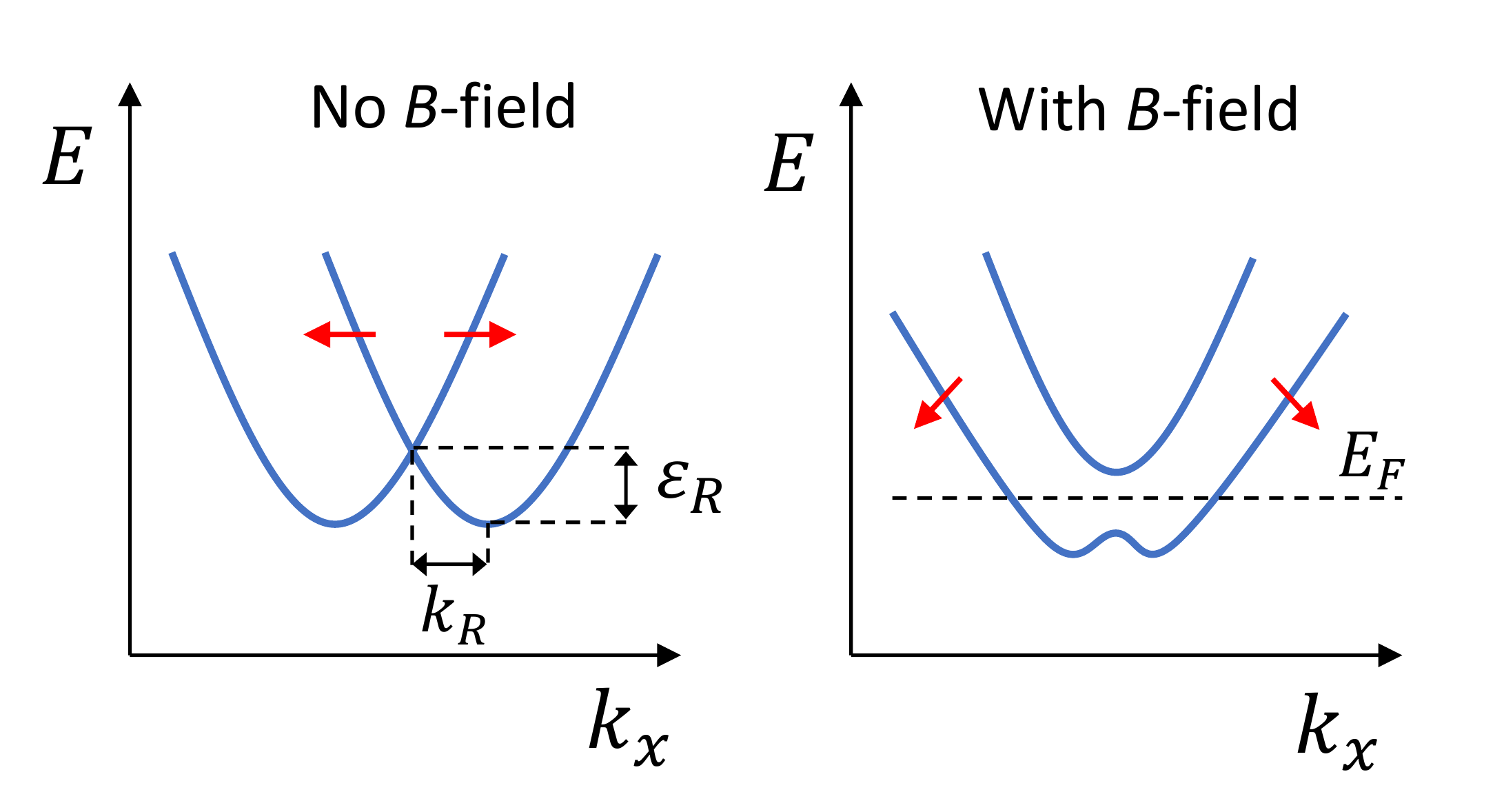}
        \caption{A spin-orbit split conduction band in the absence (left) and presence (right) of a $B$-field applied perpendicular to the spin direction. The $B$-field isolates the lower spin band leading to an effective spinless band. The Rashba wave vector and Rasba energies are indicated in the left panel. (Figure adapted from Figure 5 in Ref.~\onlinecite{majorana}).}
        \label{fig:rashba}
\end{figure}

A physical realization of the Kitaev model consists of a 1D low-density semiconducting wire with strong spin-orbit interaction (SOI) that is proximity-coupled to an $s$-wave superconductor and exposed to an external magnetic field\cite{oreg2010helical,lutchyn2010majorana}. Furthermore, the SOI should lead to a splitting of the conduction or valence band, see Fig. \ref{fig:rashba}. With the magnetic field applied perpendicular to the spin polarisation of the spin-split bands, the Zeeman effect will open a gap at the band crossing and thereby separate the lower spin band from the higher spin band.  The semiconducting wire should subsequently be gated or doped to place the Fermi level inside the lowest spin band (for electron doping) or highest spin band (for hole doping)\footnote{Note that for the $s$-wave pairing field to couple electrons of the single spin band, it is essential that the spins are not perfectly parallel throughout the 1D Brillouin zone (BZ) as the $s$-wave pairing strength scales with the size of the anti-parallel spin components.}. The setup has been intensively explored in the form of III-V semiconductor wires, most notably InAs and InSb zincblende nanowires, coupled to conventional superconducting metals like Al or Nb\cite{mourik2012signatures, das2012zero, deng2016majorana}. Here we shall not go into these fascinating experiments in any detail but refer the interested reader to the literature\cite{mourik2012signatures, deng2016majorana, de2018electric, flensberg_engineered_2021} and just mention that the question of whether true MBSs have been realized in such systems remains a topic of intense debate\cite{junger2020magnetic, valentini2021nontopological}. 

The stability of MBSs in the proximity coupled semiconductor setup, depends greatly on the size of the SOI-induced splitting of the band. The spin-splitting defines the maximum size of the energy gap between the lower and higher spin bands that is achievable by application of a magnetic field. A larger band gap enlarges the spinless regime and makes it easier to adjust the chemical potential to the right position. Moreover, a stringent condition for realising MBSs is that the spin gap exceeds the thermal energy $k_B T$. The identification of 1D materials with large SOI-induced band splitting is therefore essential for realising MBSs. Unfortunately, the spin splitting is small in the bulk form of the III-V zincblende semiconductors\cite{dresselhaus1955spin} and the realization of the MBSs thus relies on the externally induced Rashba splitting\cite{bychkov1984properties}.    

\begin{table}[tbh]
\begin{tabular}{|l|r|r|r|r|r|r|}
 \hline
 \multicolumn{1}{|c|}{Property} &
 \multicolumn{6}{|c|}{Material} \\
 \hline
& \ch{ITe}&\ch{BrTe}&\ch{ClTe}&\ch{SnClI}&\ch{SnBrCl}&\ch{SnBrI}\\ \hline

$E_\mathrm{sep}$ (eV/atom)&0.24&0.22&0.21&0.29&0.28&0.29\\ \hline
$\Delta H_\mathrm{hull}$ (eV/atom)&0.24&0.23&0.23&0.29&0.28&0.29\\ \hline
$s_\mathrm{1}$ & 0.41&0.46&0.36&0.53&0.50&0.55\\ \hline
$k_\mathrm{R}$(\AA$^{-1}$)&0.23&0.23&0.30&0.21&0.30&0.20\\ \hline
$m^\ast$ ($m_e$)&0.51&0.67&0.90&1.23&1.88&1.40\\ \hline
$\alpha $(eV \AA)&3.44&2.62&2.54&1.30&1.22&1.09\\ \hline
$\varepsilon_R^\mathrm{(eff.mass)}$ (meV)&	390&370&380&130&180&110\\ \hline

$\varepsilon_\mathrm{R}$ (meV)&100&95&95&115&120&100\\ \hline
\end{tabular}
\caption{A list of selected 1D materials from both the core and the shell that display significant Rashba splitting around the band extrema. The separation energy, $E_\mathrm{seps}$ is shown as well as the energy above the convex hull, $\Delta H_\mathrm{hull}$, calculated with PBE-D3, and the 1D scoring parameter, $s_1$. The remaining parameters, which characterize the Rashba splitting, are described in the text.}\label{tab:majorana_table}
\end{table} 

\begin{figure*}
 \centering
 \includegraphics[width=0.22\textwidth]{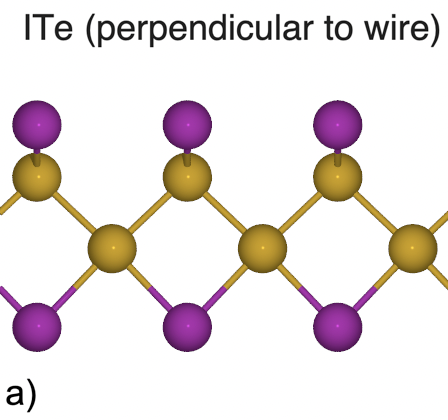}
\includegraphics[width=0.25\textwidth]{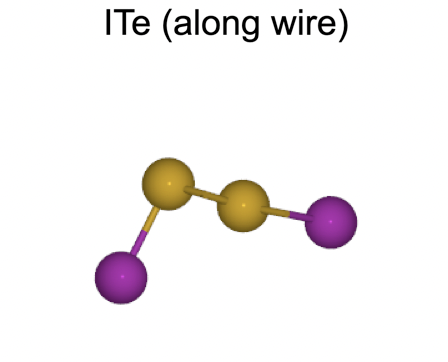}

 \includegraphics[width=0.25\textwidth]{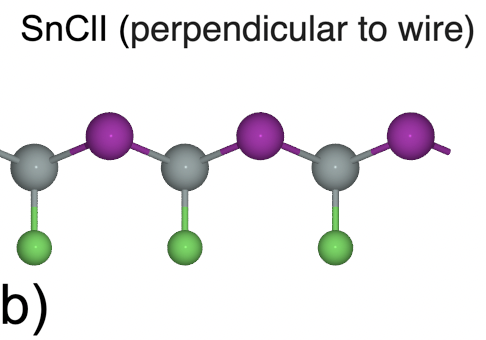}
 \includegraphics[width=0.25\textwidth]{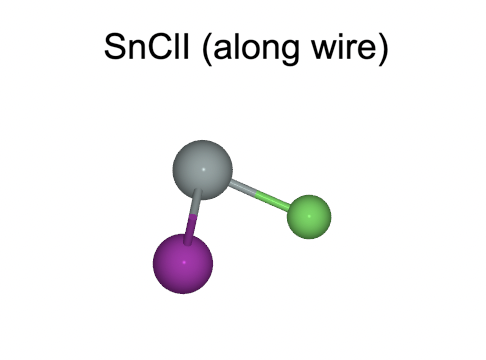}
 \caption{Crystal structures of the ITe and SnClI wires along the wire axis (left) and cross section (right). 
}
 \label{fig:ITe}
 \end{figure*}

\begin{figure*}
 \centering
 \includegraphics[width=0.3\textwidth]{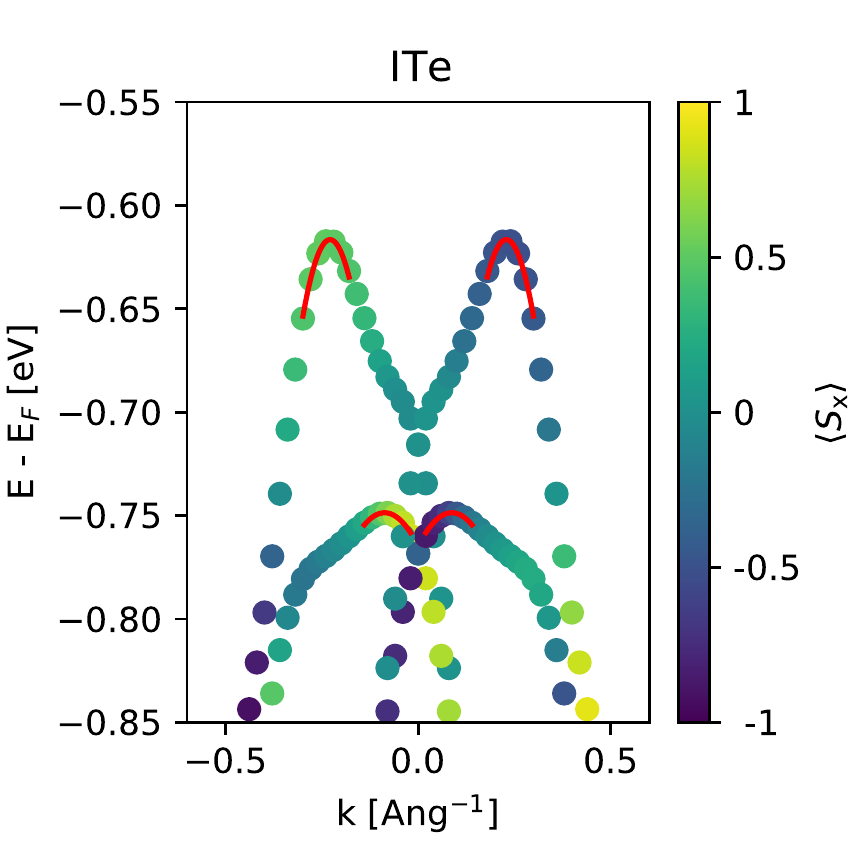}
 \includegraphics[width=0.3\textwidth]{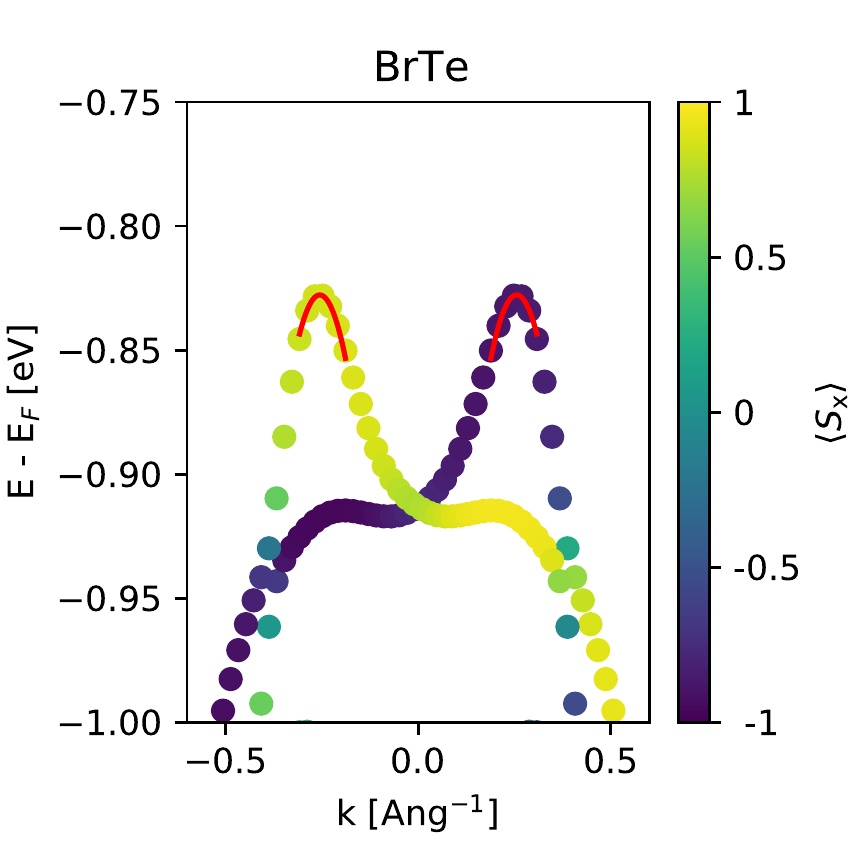}
 \includegraphics[width=0.3\textwidth]{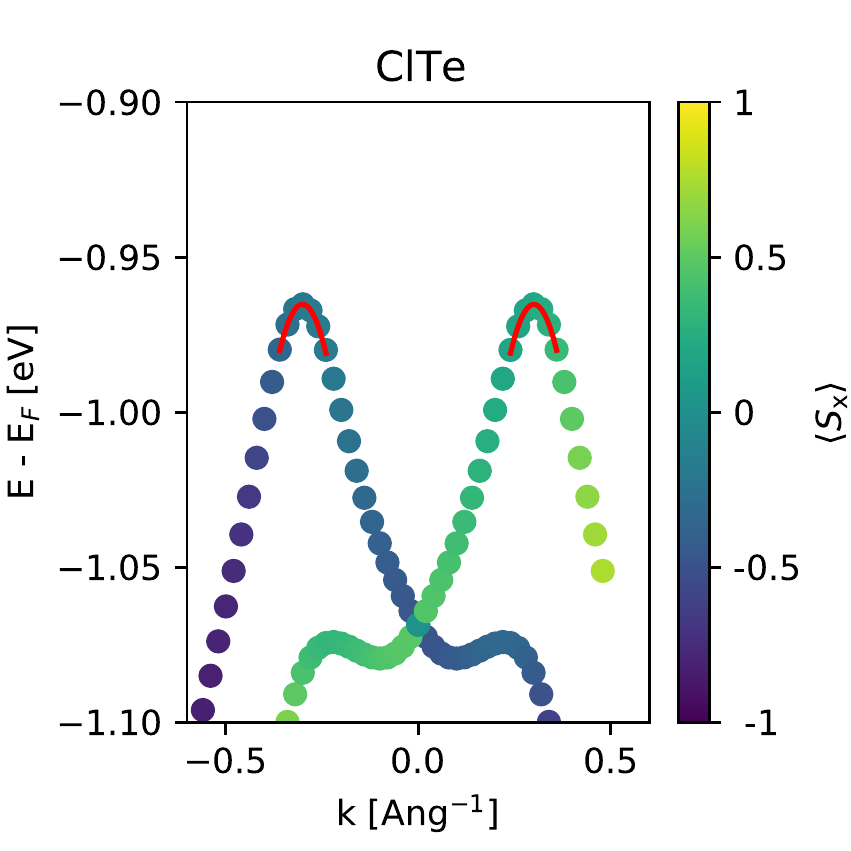}
 \includegraphics[width=0.3\textwidth]{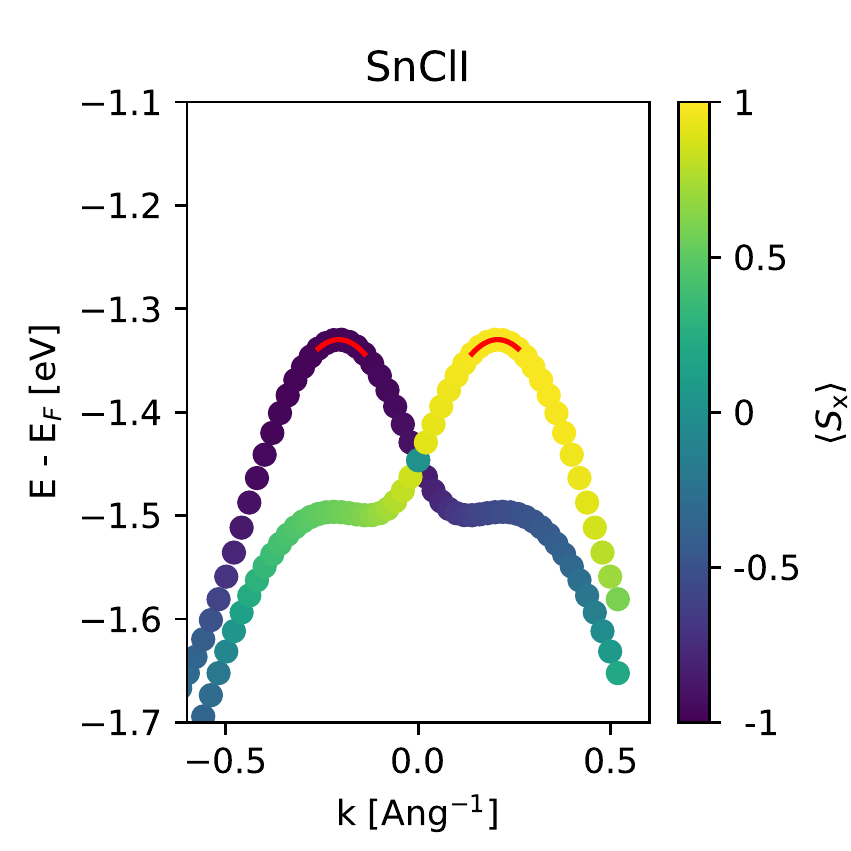}
 \includegraphics[width=0.3\textwidth]{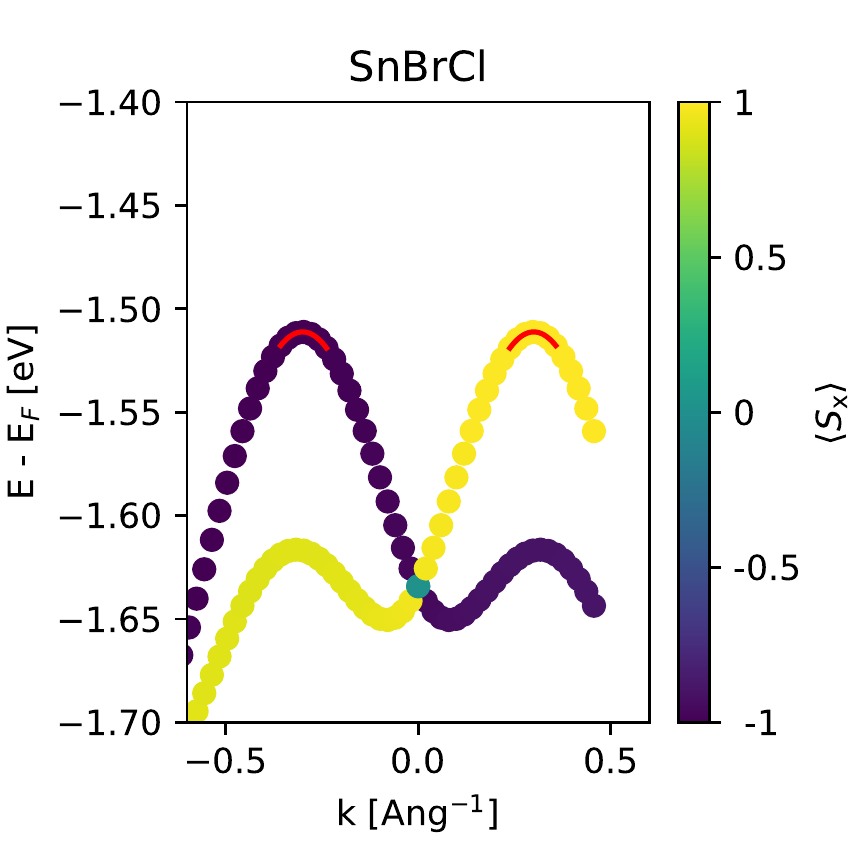}
 \includegraphics[width=0.3\textwidth]{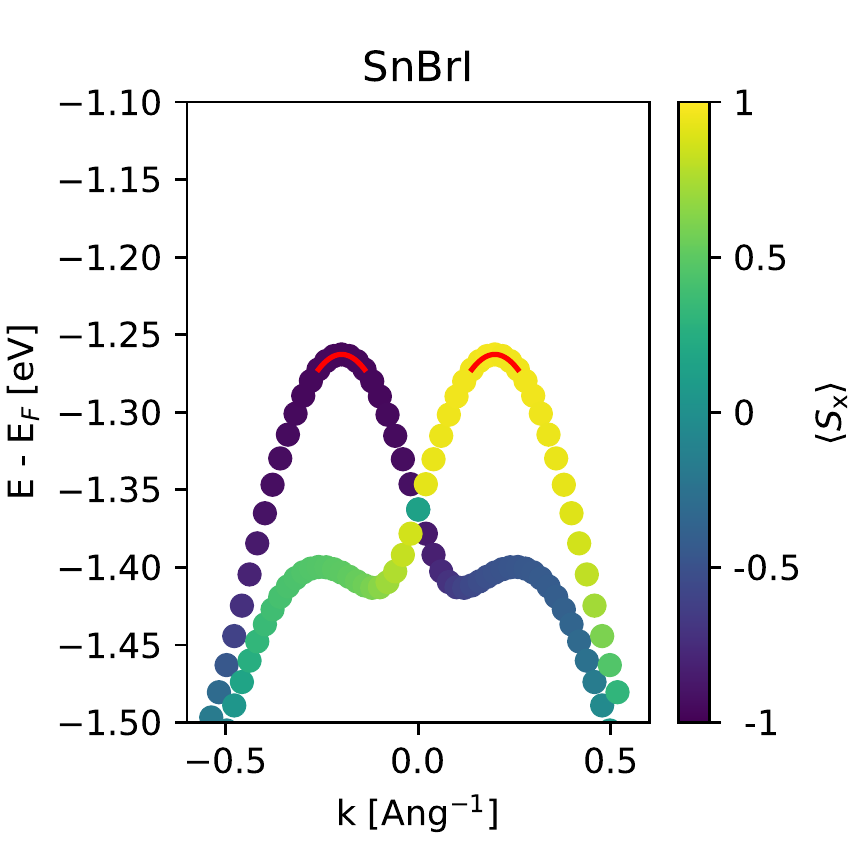}
 \caption{The band structure around the valence band maximum for the six 1D structures of the database with the largest Rashba splittings (see also Table \ref{tab:majorana_table}). The color of the bands represent the projection of the spin perpendicular to the wire axis. 
}
 \label{fig:Rashba_plot}
 \end{figure*}

Here we propose that atomically thin 1D wires, either grown "bottom-up" on a substrate or exfoliated "top-down" from bulk vdW 1D crystals, could be interesting candidates for realizing MBSs. To evaluate their potential we have performed a systematic analysis of the band structure of all the 1D structures in our database and identified the ones with largest SOI splitting. Table \ref{tab:majorana_table} lists a selection of the identified 1D structures. The listed wires all adopt one of two different crystal structures, see Fig. \ref{fig:ITe}. The selected materials have, in addition to a large spin splitting, also reasonably good stability properties with energies above the convex hull, $\Delta H_\mathrm{hull}$, in the range 0.2-0.3 eV/atom. In all cases, the largest spin splitting is found for the valence band, and the band structures close to the energy maximum are shown in Figure \ref{fig:Rashba_plot}. 
It is important to note that for the 1D structures in Table \ref{tab:majorana_table}, the SOI splitting of the bands occurs due to intrinsically broken inversion symmetry. The large size of the spin splitting (denoted $\varepsilon_R$) thus implies that there is no need to reduce the symmetry externally to enhance the splitting as required for most of the III-V nanowire systems. Despite of this, we shall stick to the conventional nomenclature and refer to the SOI-induced splitting in the 1D chains as a Rashba splitting. 

In the simplest effective-mass model, the Rashba splitting is characterized by a band structure of the form
\begin{equation}
    \varepsilon_k = -\frac{\hbar^2 k^2}{2m^\ast} + \sigma \alpha_R k,
    \label{eq:Rashba_k}
\end{equation}
where $\sigma = \pm 1$, $m^\ast$ is the effective band mass, and $\alpha_R$ is the Rashba parameter. In this model the maxima occurs at $k = \pm k_R = \pm \alpha_R m^\ast/\hbar^2$. Determining the effective mass and $k_R$ from the PBE band structure, we can thus estimate the Rashba parameter as $\alpha_R = \hbar^2k_R/m^\ast$, and the energy difference, $\varepsilon_R^\mathrm{(eff.mass)}$, between the maximum point and the band crossing at $k=0$ as $\varepsilon_R^\mathrm{(eff.mass)} = \hbar^2 k_R^2/(2m^\ast)$. The effective mass approximation is not particularly good in all cases, and we therefore also determine the energy difference, $\varepsilon_R$ between the band maximum and the crossing point at $k=0$ directly from the band structure. All these quantities are listed in Table~\ref{tab:majorana_table}.

The calculated Rashba parameters range from 1.09 eV\AA (SbBrI) to 3.44 eV\AA (ITe). In comparison, the Rashba parameters measured in state of the art III-V semconductor nanowires used as the main 1D platform for realising MBSs, are 0.5-1 eV\AA (InSb)\cite{van2015spin} and 0.1-0.3 eV\AA (InAs)\cite{liang2012strong,hansen2005spin}. More importantly, the Rashba energies ($\varepsilon_R$) in the III-V nanowires are below 1 meV\cite{van2015spin,soluyanov2016optimizing}, which is two orders of magnitude smaller than those found in the atomically thin wires, see Table \ref{tab:majorana_table}. These findings suggest that atomically thin wires could be a very interesting platform for realizing MBSs. 

In addition to a large Rashba splitting, a large Lande $g$-factor is also advantageous as it reduces the magnitude of the magnetic field required to isolate a spin band, which is important as too high magnetic field could destroy the superconducting state. Finally, a realization of MBSs in the Kitaev model also entails that $p$-wave pairing can be induced in the chain, e.g. via proximity to an $s$-wave superconductor. This question is difficult to address quantitatively, but it seems likely that proximity effects in general would be stronger for atomically thin wires as compared to the conventional MBS nanowire hosts with diameters in the range of 100 nm.  

\section{Conclusions} 
The described database is systematically generated with a core of materials extracted from the experimental ICSD and COD databases, where the degree of ``1D-ness'' is evaluated with a dimensionality scoring parameter. Furthermore, the database is extended with a shell of materials through element substitution of the core materials. The database is certainly not complete in the sense that it does not include all one-dimensional material components with a reasonable stability. An obvious limitation is the system size, which is currently limited to 20 atoms in the unit cell. We are currently expanding the database with larger systems. Another current expansion is through inclusion of element substitutions with lower probability measure, and it is also planned to make different substitutions for different atoms of the same element.

The possibility of generating new stable materials computationally using machine learning is being explored very actively \cite{schmidt_recent_2019}, and new neural network architectures are being developed with the purpose of material construction in mind \cite{xie_crystal_2022}. It would be interesting to apply these techniques to one-dimensional materials. The number of two-dimensional materials predicted computationally to be (meta-)stable is rapidly growing, and the materials exhibit a large variation in their properties \cite{gjerding_recent_2021}. We expect that a similar development will take place for one-dimensional materials.

The identification of materials, which could exhibit MBS, illustrates the usefulness of the database for computational screening studies. Future expansions in both the number of materials and in the variety of calculated properties invite for more high-throughput screening studies. The continued development of the Atomic Simulation Recipes \cite{Gjerding.2021}, which also forms the basis for the two-dimensional materials database C2DB \cite{Haastrup:2018ca, gjerding_recent_2021} makes the systematic inclusion of more properties simple and ensures a high quality of the calculations.

\section{Acknowledgements}
K.W.J. and H.M. acknowledge support from the VILLUM Center for Science of Sustainable Fuels and Chemicals, which is funded by the VILLUM Fonden research grant 9455.
Furthermore, we acknowledge funding from the European Research Council (ERC) under the European Union’s Horizon 2020 research and innovation program Grant No. 773122 (LIMA) and Grant agreement No. 951786 (NOMAD CoE). K. S. T. is a Villum Investigator supported by VILLUM FONDEN (grant no. 37789).

\bibliography{refs}
\end{document}